\documentclass[11pt]{article}
\usepackage{amsmath,amsthm,latexsym,amssymb,amsfonts,epsfig, psfrag}

\makeatletter \@addtoreset{equation}{section} \makeatother

\def\be{\begin{equation}}
\def\ee{\end{equation}}
\def\ba{\begin{eqnarray}}
\def\ea{\end{eqnarray}}

\newcommand\nn{\nonumber}
\newcommand\q{\quad}
\def\Nl{{\mathchoice
{\setbox0=\hbox{$\displaystyle\rm N$}\hbox{\hbox to0pt
{\kern0.4\wd0\vrule height0.9\ht0\hss}\box0}}
{\setbox0=\hbox{$\textstyle\rm N$}\hbox{\hbox to0pt
{\kern0.4\wd0\vrule height0.9\ht0\hss}\box0}}
{\setbox0=\hbox{$\scriptstyle\rm N$}\hbox{\hbox to0pt
{\kern0.4\wd0\vrule height0.9\ht0\hss}\box0}}
{\setbox0=\hbox{$\scriptscriptstyle\rm N$}\hbox{\hbox to0pt
{\kern0.4\wd0\vrule height0.9\ht0\hss}\box0}}}}
\def\Zl{{\mathchoice
{\setbox0=\hbox{$\displaystyle\rm Z$}\hbox{\hbox to0pt
{\kern0.4\wd0\vrule height0.9\ht0\hss}\box0}}
{\setbox0=\hbox{$\textstyle\rm Z$}\hbox{\hbox to0pt
{\kern0.4\wd0\vrule height0.9\ht0\hss}\box0}}
{\setbox0=\hbox{$\scriptstyle\rm Z$}\hbox{\hbox to0pt
{\kern0.4\wd0\vrule height0.9\ht0\hss}\box0}}
{\setbox0=\hbox{$\scriptscriptstyle\rm Z$}\hbox{\hbox to0pt
{\kern0.4\wd0\vrule height0.9\ht0\hss}\box0}}}}
\def\Ql{{\mathchoice
{\setbox0=\hbox{$\displaystyle\rm Q$}\hbox{\hbox to0pt
{\kern0.4\wd0\vrule height0.9\ht0\hss}\box0}}
{\setbox0=\hbox{$\textstyle\rm Q$}\hbox{\hbox to0pt
{\kern0.4\wd0\vrule height0.9\ht0\hss}\box0}}
{\setbox0=\hbox{$\scriptstyle\rm Q$}\hbox{\hbox to0pt
{\kern0.4\wd0\vrule height0.9\ht0\hss}\box0}}
{\setbox0=\hbox{$\scriptscriptstyle\rm Q$}\hbox{\hbox to0pt
{\kern0.4\wd0\vrule height0.9\ht0\hss}\box0}}}}
\def\Rl{{\mathchoice
{\setbox0=\hbox{$\displaystyle\rm R$}\hbox{\hbox to0pt
{\kern0.4\wd0\vrule height0.9\ht0\hss}\box0}}
{\setbox0=\hbox{$\textstyle\rm R$}\hbox{\hbox to0pt
{\kern0.4\wd0\vrule height0.9\ht0\hss}\box0}}
{\setbox0=\hbox{$\scriptstyle\rm R$}\hbox{\hbox to0pt
{\kern0.4\wd0\vrule height0.9\ht0\hss}\box0}}
{\setbox0=\hbox{$\scriptscriptstyle\rm R$}\hbox{\hbox to0pt
{\kern0.4\wd0\vrule height0.9\ht0\hss}\box0}}}}
\def\Cl{{\mathchoice
{\setbox0=\hbox{$\displaystyle\rm C$}\hbox{\hbox to0pt
{\kern0.4\wd0\vrule height0.9\ht0\hss}\box0}}
{\setbox0=\hbox{$\textstyle\rm C$}\hbox{\hbox to0pt
{\kern0.4\wd0\vrule height0.9\ht0\hss}\box0}}
{\setbox0=\hbox{$\scriptstyle\rm C$}\hbox{\hbox to0pt
{\kern0.4\wd0\vrule height0.9\ht0\hss}\box0}}
{\setbox0=\hbox{$\scriptscriptstyle\rm C$}\hbox{\hbox to0pt
{\kern0.4\wd0\vrule height0.9\ht0\hss}\box0}}}}
\def\Hl{{\mathchoice
{\setbox0=\hbox{$\displaystyle\rm H$}\hbox{\hbox to0pt
{\kern0.4\wd0\vrule height0.9\ht0\hss}\box0}}
{\setbox0=\hbox{$\textstyle\rm H$}\hbox{\hbox to0pt
{\kern0.4\wd0\vrule height0.9\ht0\hss}\box0}}
{\setbox0=\hbox{$\scriptstyle\rm H$}\hbox{\hbox to0pt
{\kern0.4\wd0\vrule height0.9\ht0\hss}\box0}}
{\setbox0=\hbox{$\scriptscriptstyle\rm H$}\hbox{\hbox to0pt
{\kern0.4\wd0\vrule height0.9\ht0\hss}\box0}}}}
\def\Ol{{\mathchoice
{\setbox0=\hbox{$\displaystyle\rm O$}\hbox{\hbox to0pt
{\kern0.4\wd0\vrule height0.9\ht0\hss}\box0}}
{\setbox0=\hbox{$\textstyle\rm O$}\hbox{\hbox to0pt
{\kern0.4\wd0\vrule height0.9\ht0\hss}\box0}}
{\setbox0=\hbox{$\scriptstyle\rm O$}\hbox{\hbox to0pt
{\kern0.4\wd0\vrule height0.9\ht0\hss}\box0}}
{\setbox0=\hbox{$\scriptscriptstyle\rm O$}\hbox{\hbox to0pt
{\kern0.4\wd0\vrule height0.9\ht0\hss}\box0}}}}
\def\nn{\nonumber}

\newcommand{\eqa}{\begin{eqnarray}}
\newcommand{\neqa}{\end{eqnarray}}

\usepackage{bbm}

\def\q{{\quad}}

\title{From covariant to canonical formulations of discrete gravity}
\author{ Bianca Dittrich\footnote{e-mail address: {\tt dittrich@aei.mpg.de}}  $^1$ and Philipp A H\"ohn\footnote{e-mail address: {\tt p.a.hohn@uu.nl}}  $^{2,3}$ \\
\small  $^1$ MPI f.~Gravitational Physics, Albert Einstein Institute,\\
 \small Am M\"uhlenberg 1, D-14476 Potsdam, Germany\\
\small   $^2$ Institute f. Theoretical Physics, Universiteit Utrecht,\\
 \small  Leuvenlaan 4, NL-3584 CE Utrecht, The Netherlands \\
\small $^3$ Institute f.~Grav.~and the Cosmos,
  The Pennsylvania State University,\\
\small 104 Davey Lab, University Park, PA 16802, USA\\
}

\date{\small AEI-2009-116 \\
ITP-UU-09/57\\
SPIN-09/47\\
IGC-09/12-1}

\begin{document}

\maketitle

\begin{abstract}
Starting from an action for discretized gravity we derive a canonical formalism that exactly reproduces the dynamics and (broken) symmetries of the covariant formalism. For linearized Regge calculus on a flat background -- which exhibits exact gauge symmetries -- we derive local and first class constraints for arbitrary triangulated Cauchy surfaces. These constraints have a clear geometric interpretation and are a first step towards obtaining anomaly--free constraint algebras for canonical lattice gravity. Taking higher order dynamics into account the symmetries of the action are broken. This results in consistency conditions on the background gauge parameters arising from the lowest non--linear equations of motion. In the canonical framework the constraints to quadratic order turn out to depend on the background gauge parameters and are therefore pseudo constraints. 

These considerations are important for connecting path integral and canonical quantizations of gravity, in particular if one attempts a perturbative expansion.
\end{abstract}

\section{Introduction}

Many approaches to quantum gravity, in particular path integral approaches such as Regge calculus \cite{regge}, introduce an auxilary discretization as a regulator into the models. Although this facilitates the construction and finiteness of the models, these methods jeopardize diffeomorphism symmetry, which is the continuum symmetry of general relativity.  This symmetry is deeply entangled with the dynamics of the theory, therefore it might be fruitful to preserve a notion of diffeomorphism symmetry as far as possible. This could largely constrain possible quantizations of discrete models, moreover, it could provide a tool for controlling lattice effects, or, in other words, the independence from the chosen discretization (usually a triangulation), see for instance the discussion in \cite{bahrdittrich09b}.  Taking care of diffeomorphism symmetry can help to obtain the correct semi--classical limit and, in particular, to obtain the correct degrees of freedom in the large scale limit (as breaking of gauge symmetries introduces additional degrees of freedom). 

There has been some discussion in the literature \cite{hamberwilliamsgauge, miller, morse, dittrich08} whether there exists an exact or only approximate notion of diffeomorphism invariance, or more generally gauge symmetry, in discretized gravity, in particular  in (the covariant form of) Regge calculus. For flat solutions it has been known \cite{hamberwilliams3d, rocekwilliams, dittrichfreidelspeziale} that symmetries exist: any vertex in the bulk of the triangulation can be translated in the embedding flat space without changing flatness, and therefore without leaving the space of solutions to the Regge equations. For a regular lattice these symmetries can be connected to the gauge modes in linearized gravity.

But in \cite{bahrdittrich09a} it was shown with the help of an explicit example that  these gauge symmetries are broken for Regge solutions with curvature.\footnote{More correctly, one expects the translation symmetry to be broken for vertices adjacent to triangles which have a non--vanishing deficit angle, that is curvature. But also in solutions with curvature there might exist vertices, for which all the deficit angles at the adjacent triangles vanish. For these the translation symmetries are preserved.} One might  take the conclusion that generically in Regge calculus there are no symmetries and that the symmetries on flat space are an exception, not very relevant for the framework. 
%However, the limit of flat space means small deficit angles, and these should also govern the continuum limit (as the deficit angles give the curvature per simplex). 
As shown in \cite{rocekwilliams}, however, the symmetries on the flat background are essential for obtaining the correct number of degrees of freedom in the continuum limit. Thus, assuming that generically there are no gauge symmetries and hence that all the degrees of freedom are physical would lead to a quite singular understanding of the continuum limit.  Indeed, in toy examples one can show that the pseudo gauge modes behave dynamically very differently from the true physical modes, having a vanishing kinematical term \cite{proceeding}. Also in  the Regge action expanded on a flat background the pseudo gauge modes obtain only a non--vanishing contribution from the (higher than second order) potential terms.

We therefore rather speak of a symmetry breaking induced by discretization and, very importantly, by the choice of the discretized action. In fact, for 3d gravity with a cosmological constant the standard Regge action leads to symmetry breaking, whereas the so--called perfect action, corresponding to choosing simplices with homogeneous curvature instead of flat ones for the discretization, does not \cite{bahrdittrich09a}. 

The breaking of symmetries has particularly severe implications for the canonical formalism. Gauge symmetries of the action lead to constraints in the canonical formalism. These are the generators of the gauge transformations and hence form an algebra which is first class.  The breaking of symmetries should therefore affect the constraints in some way.

Indeed, performing the canonical analysis of the continuum action first and then discretizing the resulting constraints leads to  additional terms that convert the continuum first class constraint algebra  into a second class constraint algebra\footnote{In frameworks where discretization is part of the regularization of quantum constraint operators, this will typically lead to anomalies in the resulting quantum algebra. Hence, one can refer to this phenomenon as classical anomalies induced by discretization or discretization anomalies.}, see for instance \cite{friedmann, loll}. Note that this approach usually involves a change of set--up for Regge calculus, namely a continuous time, but still a discretized space \cite{friedmann}.  An second class constraint algebra means that the constraints are not automatically preserved under time evolution. On the classical level one could deal with this issue by fixing the gauge parameters, that is lapse and shift, so that the constraints are preserved by time evolution \cite{friedmann,gambini}. The situation is, however, much more complicated in the quantum theory, see for instance \cite{dittrich08,lollreview} and references therein.

Nevertheless, there are attempts to derive a set of first class constraints  for discretized theories. So far these succeeded only for 3d gravity, where the theory is topological, or for 4d gravity in the sector where the triangulation is such that only flat space solutions arise \cite{zapata, dittrichryan}.  As the example of 3d gravity with a cosmological constant \cite{bahrdittrich09a} shows, the breaking of symmetries, and therefore the appearance of discretization anomalies, is not per se bound to discretization, but depends on the choice of the discrete action and, as a result, on the discrete dynamics.  There might also exist a choice of discretized constraints in the canonical framework which are first class.  In fact, in  \cite{bahrdittrich09a} it was shown that once a discrete action with symmetries has been found one can derive first class constraints. 

To use this result, one has to develop a canonical formalism which exactly reproduces the dynamics defined by the action. This means in particular to allow for a discrete time evolution. (A continuous time evolution might be recovered as a symmetry, namely the translation of vertices in time direction.) For a discretization based on triangulations such a dynamics has been proposed in \cite{dittrich08,bahrdittrich09a} based on ideas from the consistent discretization program \cite{gambini} and on the so--called tent moves \cite{commi}, which implement discrete time evolution for triangulated manifolds.  As a first step we extend these ideas to obtain a canonical description of Regge dynamics. However, we cannot expect to find exact constraints in the full theory, as the symmetries are broken for curved solutions. Nevertheless, the existence of symmetries for flat solutions should have certain repercussions. 

In order to explore these issues, we can consider an expansion of the action on a flat background. In this way the calculations become analytically tractable -- the full equations of motions are so far only solvable by numerical methods.

The linearized theory, i.e.\ the theory defined by the expansion of the action to quadratic order, has exact symmetries resulting from the null modes of the Hessian on a flat background. Hence, the canonical framework for the linearized theory should have exact constraints. We will derive these constraints explicitly for an arbitrary 3d triangulation (embeddable into 4d flat space), representing the initial data hypersurface.  One can show that these constraints are Abelian, therefore anomaly free and preserved under time evolution as defined by the tent moves. The constraints involve the background geometry in quite a complicated way. Presumably these would be hard to rederive by discretizing the continuum constraints directly. Nevertheless, these constraints are Abelian, which again follows from deriving them directly from the action but is not immediately straightforward to see on the canonical level. Moreover, deriving the canonical framework directly from the discretized action gives the momenta an immediate geometrical meaning in terms of the discrete geometry.

For the higher order dynamics one would expect that the symmetries of the action are broken, as this is the case for the full dynamics. Indeed, the equations of motion expanded to the lowest non--linear order result in consistency conditions on the background gauge, which in the case of Regge calculus is associated to the positions of the inner vertices in the flat background solution. In other words, a consistent expansion of the solutions (analytically in the expansion parameter)  is only possible for specific choices of these background gauge parameters. 

The consistency conditions on the background gauge can be rewritten as the condition that the quadratic order of Hamilton's principal function (i.e.\ the quadratic action evaluated on the solutions of the linearized theory) has a vanishing derivative with respect to the background gauge parameters. In this sense the discretization is fixed such that the dynamics depends minimally on the details of the discretization, in this case the choice of background gauge parameters.

In the canonical framework it will turn out that the quadratic order of the constraints depends on the background gauge parameter. These could be interpreted as background lapse and shift and hence we encounter (lapse and shift dependent) pseudo constraints rather than exact constraints. The requirement that the constraints should be preserved under the discrete time evolution leads to the same condition on the background gauge parameter as in the covariant formalism.

\vspace{0.3cm}

The structure of the paper is as follows:  
In section  \ref{summary} we will give a short introduction to Regge calculus and its equations of motion. Subsequently, in section \ref{can}, we will discuss the tent moves and the associated canonical formalism. In section \ref{bianchi} we explain the origin of the gauge symmetries for flat solutions and the relation to the Bianchi identities. The symmetries of the flat solution imply that the Hessian of the action evaluated on these solutions will have null vectors, which we will examine in section \ref{deg}. These considerations are essential in order to derive the constraints for the linearized theory in section \ref{lin}. In section \ref{4valent} and section \ref{sechigher} we will detail the constraints for four-- and higher valent vertices. We show in section \ref{algebra} that the constraints are Abelian and, after performing a split into linearized observables and gauge variables in section \ref{obs}, we consider the dynamics of the linearized observables -- that is gravitons -- as generated by the tent moves in section \ref{dyn}. In particular, we will show that the constraints are consistent, that is automatically preserved under time evolution. Finally, we discuss the higher order dynamics in section \ref{higher}. We will close with a summary and outlook.

\section{Summary of Regge calculus} \label{summary}

Regge calculus \cite{regge} is usually considered on a fixed triangulation $\mathcal{T}$ of a space--time manifold. We will consider 4d triangulations which are built from 4--simplices. A 4--simplex has five tetrahedra $\tau$, ten triangles $\Delta$, ten edges $e$ and five vertices $v$ as subsimplices. Two 4--simplices are glued together by identifying a tetrahedron from each 4--simplex with each other. Thus, a tetrahedron (in the bulk of the triangulation) is always shared by two 4--simplices, whereas, for instance, a triangle can be shared by any number (higher than two) of 4--simplices. 

The variables appearing in the Regge action, which defines the equations of motion for Regge calculus, are usually given by the edge lengths $\{l^e\}_{e \in \mathcal{T}}$, for other choices see \cite{dittrichspeziale,dittrichbahr09c}.  These variables completely specify the (piecewise linear) geometry of the triangulation.  In particular, from the edge lengths one can compute any 4d dihedral angle $\theta^\sigma_\Delta$, which give the (inner) angle in the 4--simplex $\sigma$ between the two tetrahedra sharing the triangle $\Delta$. These dihedral angles in turn determine the curvature of the triangulation:

Consider a triangle $\Delta$ in a 4d triangulation. This triangle is shared by several 4--simplices. A (Levi--Civita) parallel transport of a vector from one 4--simplex to the next around the triangle results in a rotation of this vector by the so--called deficit angle $\epsilon_\Delta=2\pi-\sum_{ \sigma \supset \Delta} \theta^\sigma_\Delta$ in the plane perpendicular to this triangle. The deficit angle measures the curvature concentrated at the triangle. 

Accordingly, the (Euclidean) Regge action, as a discretization of the Einstein--Hilbert action\footnote{We work in units with $c=8\pi G_{Newton}=1$.} $S_{EH}=-\tfrac{1}{2}\int \sqrt{g}R d^4x$,  is given by
\be\label{reggeaction1}
S=-\sum_{\Delta \subset\text{bulk}} A_\Delta \epsilon_\Delta + S_{bdry} \q ,
\ee
where $A_\Delta$ denotes the area of a triangle $\Delta$. If there is a non--vanishing boundary the boundary term 
\be\label{bdry}
S_{\text{bdry}}=-\sum_{\Delta \subset \text{bdry}} A_\Delta \psi_\Delta \q 
\ee
has to be added to the action in order to make the boundary value problem (with prescribed edge lengths on the boundary) well defined. Here $\psi_\Delta=k\pi-\sum_{ \sigma \supset \Delta} \theta^\sigma_\Delta$ is the extrinsic curvature angle. The value $k$ is determined by how many pieces are glued together at the triangle $\Delta$ in question. Usually only two pieces are added, in which case $k=1$ to ensure that the actions for the two pieces add up correctly to the action for the glued triangulation. Sometimes more pieces are glued together, in this case we will use $k=0$ for the additional pieces.

To obtain the equations of motion one has to vary the Regge action (\ref{reggeaction1}) with respect to the length variables. Here the Schl\"afli identity is instrumental. The Schl\"afli identity 
\ba\label{schlaefli}
\sum_{\Delta \subset \sigma} A_\Delta \delta \theta^\sigma_\Delta=0
\ea
relates the variatons $\delta \theta^\sigma_\Delta$ of the dihedral angles in a 4--simplex\footnote{This identity can be generalized to any $n$--dimensional simplex.}.

 The Schl\"afli identity can be understood to be analogous to the result that for the variation of the  Einstein--Hilbert action $-\int \sqrt{g} g^{ab}R_{ab} d^4x$ the term with the variation of the Ricci tensor leads to a total divergence. Indeed, also for the Regge action the variations of the deficit angles lead to a contribution only from the boundary, which are annihilated by the variation of the boundary term (\ref{bdry}).  The resulting equation of motion obtained by varying the length of the edge $e$ in the bulk is
 \ba\label{eom0}
 -\sum_{\Delta \supset e}    \frac{\partial A_\Delta}{\partial l^e} \epsilon_\Delta =0  \q .
 \ea
A special kind of solutions are flat triangulations, for which all deficit angles vanish, $\epsilon_\Delta=0$. Flat solutions, however, can only appear for specific choices of the boundary lengths (if the triangulation of the boundary is sufficiently complicated\footnote{There are special (simple) types of boundary triangulations, for instance the boundary of a 4--simplex, for which flat solutions are generically possible, i.e.\ for generic choices of the boundary lengths. The reason is that these 3d triangulations can always be embedded into 4d flat space.}).

Note that if the boundary lengths allow for a flat solution and the triangulation contains vertices in the bulk, this solution is not unique. That is, other flat solutions can be produced by translating the inner vertices in the embedding 4d flat space and changing the lengths of the inner edges adjacent to these vertices accordingly. On the other hand, one expects that for boundary data inducing curvature, the solutions are unique. A family of examples was considered numerically in \cite{bahrdittrich09a} and uniqueness of the solutions was found.

\section{Canonical formalism}\label{can}

As mentioned, the edge lengths are not uniquely determined for boundary conditions such that the Regge equations admit flat solutions. There is rather a gauge freedom, which can be understood from the choice of the exact position of the inner vertices in the flat triangulation. Such a gauge freedom is usually accompanied by constraints in the canonical formulation of the theory. Here, however, we encounter the situation that only a particular set of solutions, namely the flat ones, exhibits an exact form of gauge freedom. We will show in section \ref{lin} that this leads to constraints in the canonical formulation of  linearized  theories around  flat solutions. Later on, we will discuss the repercussions for the full dynamics of the canonical theory. 

To define the canonical formulation along the lines of  \cite{dittrich08, bahrdittrich09a} we will employ the so--called tent moves, introduced in \cite{commi}. These admit the advantage to allow for a local evolution of the hypersurface on which the canonical data are defined, without changing the connectivity of its triangulation. Therefore the number of edges and hence variables remains constant under the discrete time evolution. Note that although we use the tent moves to derive the (linearized) constraints it will turn out that these are independent from this construction.  The constraints are conditions on the canonical data, so that these can be consistently evolved. Alternatively, (by evolving backwards) one can see the constraints as describing canonical data that can arise by an evolution leading to the hypersurface in question.
 
\begin{figure}[hbt!]
\begin{center}
    \psfrag{vn}{$v_{n}$}
    \psfrag{vn+1}{$v_{n+1}$}
    \psfrag{1}{$1$}
    \psfrag{2}{$2$}
    \psfrag{3}{$3$}
    \psfrag{4}{$4$}
    \includegraphics[scale=0.3]{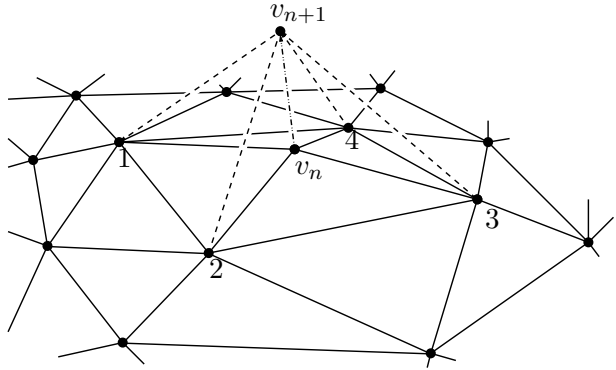}
    \end{center}
    \caption{\label{3dtm1}{\small The tent move in 3d applied to a vertex $v_n$ in a 2d Cauchy hypersurface.}}
\end{figure}

\begin{figure}[hbt!]
\begin{center}
    \psfrag{vn}{$v_{n}$}
    \psfrag{vn+1}{$v_{n+1}$}
    \psfrag{1}{$1$}
    \psfrag{2}{$2$}
    \psfrag{3}{$3$}
    \psfrag{4}{$4$}
    \includegraphics[scale=0.4]{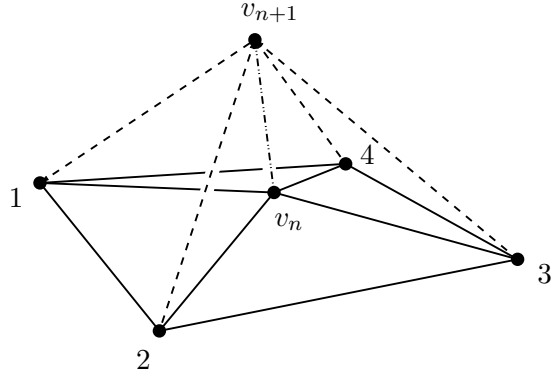}
    \end{center}
    \caption{\label{3dtm2}{\small The local tent move evolution of a vertex $v_n$ in 3d.}}
\end{figure}

To define a tent move
consider a 3d triangulation $\Sigma_n$, which can be
thought of as a triangulated Cauchy hypersurface with time label $n$. We will assume that this Cauchy hypersurface is a (piece of a) boundary of a 4d triangulation, whose inner edge lengths satisfy the Regge equations. Pick a vertex $v_n$
in the Cauchy surface and define a new vertex $v_{n+1}$, which will lie in the evolved Cauchy hypersurface $\Sigma_{n+1}$. Connect both vertices with an edge, which will be called `tent pole'. 
Denote all
other vertices in $\Sigma_n$ which $v_n$ is connected to by $1,\ldots,
N$. Connect also $v_{n+1}$ to the $1,\ldots, N$ by edges. Furthermore, we
will have a tetrahedron $\tau(v_{n+1}ijk)$  (with vertices $v_{n+1},i,j,k$) in the evolved hypersurface $\Sigma_{n+1}$ for
every tetrahedron $\tau(v_nijk)$ in $\Sigma_n$. Hence, the triangulations of the
two Cauchy surfaces are the same. The analogous evolution of a tent move in 3d is depicted in figures \ref{3dtm1} and \ref{3dtm2}.

\noindent

The evolution can be thought of as
gluing a certain piece of 4--dimensional triangulation onto the
hypersurface. This 4--dimensional triangulation $\mathcal{T}_n$ consists of
4--simplices $\sigma(v_nv_{n+1}ijk)$ for every tetrahedron $\tau(v_nijk)$ in $\Sigma_n$. Note that the tent move only involves $\text{star}(v_n)$, the (3d) star of $v_n$ in $\Sigma_n$. The star of a vertex is the union of all simplices having $v$ as a subsimplex. 

Through this gluing--on we have obtained an additional $(N+1)$ inner edges, namely the edges $e=e(v_ni),i=1,\ldots,N$ and the tent pole $t=e(v_n v_{n+1})$. We will rewrite the equations of motion for these edges into a canonical form. To this end denote by 
\ba\label{oct1}
S_n &=&
-\sum_{\Delta \subset \overset{\circ}{\mathcal T}_n} 
A_\Delta \left( 2\pi-\sum_{\sigma\subset {\mathcal T}_n}       \theta^\sigma_{\Delta}\right)
\nn\\
&& -\sum_{\Delta \subset \overset{\circ}{\text{star}(v_n)}} A_{\Delta }
  \left(\pi-\sum_{\sigma\subset {\mathcal T}_n}       \theta^\sigma_{\Delta}\right)
  -\sum_{\Delta_{}\subset\overset{\circ}{\text{star}(v_{n+1})}} A_{\Delta} 
  \left(\pi-\sum_{\sigma\subset{\mathcal T}_n }       \theta^\sigma_{\Delta}\right) \nn\\
 && -\sum_{\Delta \subset \text{star}(v_n) \cap \text{star}(v_{n+1})} A_\Delta\left(  -\sum_{\sigma \subset {\mathcal T}_n} \theta^\sigma_\Delta \right)
\ea
the Regge action for the added piece of 4d triangulation $\mathcal{T}_n$ (with boundary terms).  With $\Delta \in \overset{\circ}{\text{star}(v_n)}$  (or $\Delta \in \overset{\circ} {\text{star}(v_{n+1})}$) we mean triangles that are in $\text{star}(v_n)$ but are not part of $\text{star}(v_{n+1})$ (or vice versa).  There are also triangles which are part of both $\Sigma_n$ and $\Sigma_{n+1}$. If one performs several consecutive tent moves at the vertices $v_n,v_{n+1},\dotsc$, then these triangles are part of each of the triangulations $\mathcal{T}_n,\mathcal{T}_{n+1},\dotsc$. Hence, we choose the associated boundary term without any factor of $\pi$, as we cannot say how many pieces $\mathcal{T}$ are added. (Also if tent moves at neighboring vertices are performed then the action associated to these moves provides the necessary factors of $\pi$ for these triangles.)

With $S_{n-1}$ we will denote the action (again with boundary terms) of the original 4d triangulation without the piece $\mathcal{T}_n$. (Alternatively, one can assume that a tent move at $v_{n-1}$ has already been performed. Then $S_{n-1}$  is the action associated to $\mathcal{T}_{n-1}$. Again, this does not matter for the equations of motion.) The equations of motion can be written as
\ba\label{oct2}
0&=& \frac{\partial S_n}{\partial t_n} \nn\\
0&=& \frac{\partial S_{n-1}}{\partial l^e_n} +  \frac{\partial S_{n}}{\partial l^e_n} 
\ea 
where by $t_n$ we denote the length of the tent pole $t=e(v_n v_{n+1})$ and $l^e_n$ is the length of the edge $e=e(v_ni),i=1,\dotsc,N$. Using $S_n$ as a generating function, we define the momenta canonically conjugate to $l^e_n,l^e_{n+1},t_n,t_{n+1}$ by
\begin{xalignat}{2}\label{oct3}
&p_t^n\,\,\,:= - \frac{\partial S_n}{\partial t_n}  &&
p_e^n\,\,\,:= - \frac{\partial S_{n}}{\partial l^e_n} \nn\\
&p_t^{n+1}:=  \frac{\partial S_n}{\partial t_{n+1}}   &&
p_e^{n+1}:=  \frac{\partial S_{n}}{\partial l^e_{n+1}}   \q .
\end{xalignat}
Note that the momentum $p^{n+1}_t$ identically vanishes as $S_n$ does not depend on $t_{n+1}$.
The equations of motion (\ref{oct2}) are now simply given by
\ba\label{oct4}
p^n_t&=&\frac{\partial S_{n-1}}{\partial t_n}=-\frac{\partial S_{n}}{\partial t_n}=p^n_t =0\nn\\
p^n_{e}&=&\frac{\partial S_{n-1}}{\partial l^e_n}=-\frac{\partial S_{n}}{\partial l^e_n}=p^n_e  \q ,
\ea
and thus reproduce the Regge equations of motion (\ref{oct2}).
%\ba\label{zus1}
%frac{\partial S}{\partial l^e_n}=0 \q , \q\q \frac{\partial S}{\partial t_n}=0  \q 
%\ea
%where $S=S_{n-1}+S_n$ contains the parts of the action that depend on $l^e_n$ or $t_n$.  

\noindent

\noindent

Apart from the edges $e$ adjacent to $v_n$ there are more edges $b$ in (the boundary of) the 3d star of $v_n$. The lengths of these edges do not change under a tent move at $v_n$, however, if one performs tent moves at neighboring vertices one has to transform the momenta associated to these edges. This transformation is dictated by the requirement to reproduce the Regge equation of motion for all tent moves. This is achieved by defining a generating function (of mixed type)
\ba\label{zus2}
F(l^e_n,l^e_{n+1},t_n, l^b_n,p^{n+1}_b)=-\sum_{b} l^b_n \, p_b^{n+1} + S_n(l^e_n,l^e_{n+1},t_n,l^b_n) \q .
\ea
The transformations for the variables associated to $t,e$ do not change from (\ref{oct3}). For the edges $b$ in the boundary of $\text{star}(v_n)$ we obtain
\ba\label{zus3}
l^b_{n+1}&:=&-\frac{\partial F}{\partial p_b^{n+1}}=l^b_n \nn\\
p^n_b&:=& -\frac{\partial F}{\partial l^b_{n}} \q= p_b^{n+1}- \frac{\partial S_{n}}{\partial l^b_n}  \q .
\ea
As we have $l^b_n=l^b_{n+1}$ we will often use just $l^b$ for these variables.

The canonical transformation (\ref{oct3}, \ref{zus3}) (it is canonical because we defined it via a generating function) defines the evolution of the canonical data from time step $n$ to time step $(n+1)$. Here the equation of motion $p_t^n=0$ is analogous to the primary constraints appearing in continuum (canonical) general relativity $p_{\mathcal{N}}=0$, which imply that the momentum conjugate to the lapse function vanishes. The momenta at time step $n$ as defined by (\ref{oct3}) are explicitly
\ba\label{oct5}
p_t^n& =& \sum_{i=1}^{N}   \frac{\partial A_{\Delta(v_n v_{n+1}i)}}{\partial t_n} \epsilon_{\Delta(v_n v_{n+1}i)}   \nn\\
p_{e}^n &=&     \sum_{\Delta \in \overset{\circ} {\text{star}(v_n)} }   \frac{\partial A_\Delta}{\partial l^e_n} \psi_{\Delta}
+\frac{\partial A_{\Delta(v^n v^{n+1}i)}}{\partial l^e_n} \epsilon_{\Delta(v_n v_{n+1}i)}
\ea
with $e=e(v_n i)$ -- the edge connecting $v_n$ and the vertex $i$.  For the new momenta we obtain
\ba\label{oct5b}
p_{e}^{n+1} &=&  -   \sum_{\Delta \in \overset{\circ} {\text{star}(v_{n+1})} }   \frac{\partial A_\Delta}{\partial l^e_{n+1}} \psi_{\Delta}
-\frac{\partial A_{\Delta(v^n v^{n+1}i)}}{\partial l^e_{n+1}} \epsilon_{\Delta(v_n v_{n+1}i)}  \nn\\
p_b^{n+1}&=&p_b^{n} -    \sum_{\Delta \in \overset{\circ}{\text{star}(v_{n})} }   \frac{\partial A_\Delta}{\partial l^b} \psi_{\Delta}
-    \sum_{\Delta \in  \overset{\circ}    {\text{star}(v_{n+1})} }   \frac{\partial A_\Delta}{\partial l^b} \psi_{\Delta}
- \sum_{\Delta \in {\text{star}(v_{n})} \cap {\text{star}(v_{n+1})}}  \frac{\partial A_\Delta}{\partial l^b} \psi_{\Delta}\,.\q\q\;
\ea

\section{Gauge symmetry and contracted Bianchi identities} \label{bianchi}

In continuum general relativity the contracted Bianchi identities,
\ba\label{bi1}
\nabla^aG_{ab}=0\q,
\ea
for the Einstein tensor $G_{ab}=R_{ab}-\tfrac{1}{2}R g_{ab}$  can be derived from the invariance of the Einstein--Hilbert action under diffeomorphisms, see for instance \cite{wipf,wald}. On the other hand, the contracted Bianchi identities are geometrical identites, which follow from the properties of the curvature tensor.

From the contracted Bianchi identities it follows that not all ten of Einstein's field equations are independent. The identities $\nabla^aG_{ab}=0$ provide four differential relations between the field equations; in other words, the field equations are not fully independent from each other. Given sufficient initial data, the evolution of the ten metric components is therefore not completely determined. This is related to the freedom of choosing coordinates and therefore ultimately to general covariance. 

Similar arguments can be made for Regge calculus \cite{miller, morse, rocekwilliams, hamber}, in which the Bianchi identities hold as geometrical identities, giving relations between finite rotation matrices \cite{hamber, freidellouapre}. There is, however, a difference to the continuum case, namely that the relation to the equations of motion can only be made approximately, only valid for small deficit angles (and further assumptions on the `fatness' of the simplices \cite{morse}). This means that the equations of motion will be dependent only in this approximation. However, this approximation turns into an exact identity for the linearized theory on a flat background.

%Since near-flat triangulations of Regge calculus should provide good approximations to solutions of GR we expect a similar relation to hold in linearized Regge calculus. We will explicitly investigate this point for the linearized tent move program. As a prerequisite we firstly explain how the gauge transformations given by $Y^e_I$ can be viewed as a result of the contracted Bianchi identity.

In the following discussion we will adapt some arguments from \cite{morse} to the linearized theory and clarify the origin of the degeneracy of the Hessian of the Regge action which we will discuss in section \ref{deg}.  In a triangulation of flat space one can displace the vertices in the embedding flat space without changing the deficit angles (which are vanishing). The induced infinitesimal change of length variables $\delta l^e$ is described by the vector fields $\delta l^e=Y^e_I$  ($I=1,\ldots ,4$) which can easily be computed to 
\be
Y^e_I=\frac{\vec{B}_I\cdot \vec{E}^e}{\sqrt { \vec{E}^e \cdot  \vec{E}^e}}
\ee
where $\vec{B}_I$ is a basis in the embedding flat 4d space and $\vec{E}^e$ are the 4d vectors for the edges $e$ adjacent to  a given vertex $v$ (all edges are either pointing towards the vertex or pointing away). For edges $e$ not adjacent to $v$ the components of $Y_I^e$ are zero. The vector fields $Y^e_I$ act trivially on the deficit angles (evaluated on flat backgrounds), that is
\ba\label{bia2}
Y^e_I  \frac{\partial \epsilon_\Delta}{\partial l^e}\,\,{}_{|flat}=0   \q .
\ea
Taking the sum over triangles and multiplying with a factor $\partial A_\Delta/\partial l^{e'}$ we obtain
\ba\label{bia3}
Y^e_I  \sum_{\Delta} \frac{\partial A_{\Delta}}{\partial l^{e'}}  \frac{\partial \epsilon_\Delta}{\partial l^e}{}_{|flat}=0 \q .
\ea

Note that the Hessian of the Regge action (with respect to the length variables of edges in the bulk) evaluated on a flat background is exactly
\ba
\frac{\partial^2 S}{\partial l^{e'}\partial l^e} =-\sum_{\Delta} \frac{\partial A_{\Delta}}{\partial l^{e'}}  \frac{\partial \epsilon_\Delta}{\partial l^e}{}_{|flat}
\ea
so that  equation (\ref{bia3}) shows that this Hessian is degenerate and that the derivatives appearing in (\ref{bia3}) commute. Expanding the lengths as $l^e={}^{(0)}l^e+ \varepsilon y^e+ O(\varepsilon^2)$ we therefore find that the equations of motion for the linearized theory are linearly dependent,
\ba\label{bia5}
{}^vY^e_I  \sum_{\Delta} \frac{\partial A_{\Delta}}{\partial l^{e}}  \frac{\partial \epsilon_\Delta}{\partial l^{e'}}{}_{|flat} \,\,y^{e'}=0 \q .
\ea

This equation is a first order expansion of the `approximate Bianchi identity' \cite{miller, morse}
\ba\label{001a}
{}^vY^e_I  \sum_{\Delta} \frac{\partial A_{\Delta}}{\partial l^{e}} \epsilon_\Delta \approx 0 \q ,
\ea
expressing that the equations of motion for the full theory are approximately dependent on each other. Conversely, equation (\ref{001a}) can be used to show that the Hessian of the Regge action is degenerate -- equation (\ref{bia3}) -- when the linearized Bianchi identity holds.

The fact that the equations of motions are not independent from each other leaves some of the variables -- four per vertex -- undetermined, which explains the gauge freedom appearing in the (linearized) theory.

\section{Degeneracies of the Hessian of the action}  \label{deg}

The linearized theory is defined by the quadratic expansion of the action around a background solution. As the linear terms vanish due to the background satisfying the equations of motions, the linearized theory is basically determined by the Hessian of the action, i.e.\ the matrix of second derivatives. Consequently, we need to understand the properties of this Hessian (here evaluated on a flat solution) in order to understand the properties of the linearized theory.

Assume that we have performed two consecutive tent moves so that the triangulation that was added in the process is $\mathcal{T}_{n-1} \cup \mathcal{T}_n$. This piece of triangulation has one inner vertex $v_{n}$. Consider this piece of triangulation as a boundary value problem, i.e.\ fix all the edge lengths in the boundary $\Sigma_{n-1} \cup \Sigma_{n+1}$ and solve the Regge equations for the $(N+2)$ inner edge lengths $\{l^e_n, e=e(v_ni),i=1,\ldots,N\}$ and $t_{n-1},t_n$. Furthermore, assume that the boundary lengths admit a flat solution, i.e.\ a solution for which all the deficit angles at the inner triangles vanish.\footnote{Such boundary conditions can be constructed as follows: Start with the piece of 3d triangulation $\Sigma_{n-1}$ and embed this into flat 4d space. Now perform the tent move but within the embedding of the flat space. That is, after choosing the position of the tip of the tent pole $v_{n+1}$ (and the lengths in $\Sigma_{n-1}$)  the lengths of the edges from the tip of the tent pole to the vertices $i=1,\ldots,N$ are determined.  This gives a boundary that - as it is embedded in flat space - admits a flat solution. The boundary problem is solved by just choosing the position of the inner vertex $v_n$ and connecting this vertex with the vertices in the boundary appropriately.} Now the triangulation  $\mathcal{T}_{n-1} \cup \mathcal{T}_n$ represents a piece of flat space and by moving the vertex around inside this piece of flat 4d space and changing the length of the adjacent edges appropriately, i.e.\ according to their embedding into flat space, we will obtain a 4--parameter set of further solutions. 

Hence, the extremum of the action corresponding to the original flat solution is not an isolated one, rather there are four constant directions at this extremum. Accordingly, and as discussed in the previous section, the Hessian of this action with respect to the inner edge lengths $\{l^e_n, e=e(v_ni),i=1,\ldots,N\}$ and $t_{n-1},t_n$ has four null vectors $Y_I,I=1,\ldots,4$, whose components we will denote by $(Y^{t_{n-1}}_I, Y^{t_n}_I, \{Y^e_I\})$. 

In the following we will show that from these null vectors we can define null vectors for other Hessians that will appear in the dynamics of the linearized theory. To begin with we will eliminate the lengths of the tent poles as these function are auxiliary variables.  To this end we define the `effective action' 
\ba\label{oct6}
\tilde S(l^e_{n-1},l^e_n,l^e_{n+1},l^b)&:=&S_{n-1}(T_{n-1}(l^e_{n-1},l^e_n,l^b),\,\,l^e_{n-1},l^e_n,l^b)\,+\,S_{n}(T_{n}(l^e_{n},l^e_{n+1},l^b),\,\,l^e_{n},l^e_{n+1},l^b) \nn\\
&=& \tilde S_{n-1}(l^e_{n-1},l^e_n,l^b)\q\q\q\q\q\q\q+\, \tilde S_n(l^e_n,l^e_{n+1},l^b)
\ea
where we have solved the associated equations of motion for the length of the tent poles, so that
\begin{xalignat}{2}\label{oct7}
&\frac{\partial S_{n-1}}{\partial t_{n-1}}(T_{n-1}(l^e_{n-1},l^e_n),\,\,l^e_{n-1},l^e_n) \equiv 0 ,&&  \frac{\partial S_{n}}{\partial t_{n}}(T_{n}(l^e_{n},l^e_{n+1}),\,\,l^e_{n},l^e_{n+1}) \equiv 0   \q .
\end{xalignat}

The following arguments will show that the $\{Y^e_I\}$ define the null vectors of the Hessian of the `effective action'.  Taking the derivative of these equations (\ref{oct7}) with respect to $l^e_n$ we obtain the identities 
\ba\label{oct8}
\frac{\partial^2 S_{n-1}}{\partial t_{n-1} \partial l_{n}^e}+ \frac{\partial^2 S_{n-1}}{\partial t_{n-1} \partial t_{n-1}} \frac{\partial T_{n-1}}{\partial l_{n}^e} &=&0 \nn\\
 \frac{\partial^2 S_{n}}{\partial t_{n} \partial  l_{n}^e}+ \frac{\partial^2 S_{n}}{\partial t_{n} \partial t_n} \frac{\partial T_{n}}{\partial l_{n}^e} &=&0  \q ,
\ea
so that
\ba\label{oct7a}
 \frac{\partial T_{n-1}}{\partial l_{n}^e} &=&- \left(   \frac{\partial^2 S_{n-1} }{\partial t_{n-1} \partial  t_{n-1}}     \right)^{-1}  \frac{\partial^2 S_{n-1} }{ \partial t^{n-1} \partial l^n_e} \nn\\
 \frac{\partial T_{n}}{\partial l_{n}^e} &=&- \left(   \frac{\partial^2 S_{n} }{\partial t_{n} \partial  t_{n}}     \right)^{-1}  \frac{\partial^2 S_{n} }{ \partial t^{n} \partial l^n_e} \q .
\ea

On the other hand, following from the fact that the $Y_I$ are null vectors of the Hessian of $S=S_{n-1}+S_n$ (we will apply the Einstein summation convention for the index $e$) we have 
\ba\label{oct9}
  Y^e_I \frac{\partial^2 S }{\partial l^e_n
\partial l^{e'}_n}
 + Y^{t_{n-1}}_I  \frac{\partial^2 S }{\partial t_{n-1} \partial l^{e'}_n}
+ Y^{t_{n}}_I  \frac{\partial^2 S }{\partial t_{n} \partial l^{e'}_n}
&=&0   \nn\\
 Y^e_I \frac{\partial^2 S }{\partial l^n_e \partial t^{n-1}}
 + Y^{t_{n-1}}_I  \frac{\partial^2 S }{\partial t_{n-1} \partial  t_{n-1}}
&=&0   \nn\\
 Y^e_I \frac{\partial^2 S }{\partial l_n^e \partial t_{n}}
 + Y^{t_{n}}_I  \frac{\partial^2 S }{\partial t_{n}  \partial t_{n}}
&=&0     \q .
\ea
From the last two equations in (\ref{oct9}) we obtain the components $Y_I^{t_{n-1}}$ and $Y_I^{t_{n}}$ as functions of $Y^{e}_I$
\ba\label{oct9a}
Y^{t_{n-1}}_I &=&  -Y^e_I \frac{\partial^2 S_{n-1} }{\partial l^n_e \partial t^{n-1}}\left(   \frac{\partial^2 S_{n-1} }{\partial t_{n-1} \partial  t_{n-1}}     \right)^{-1}  \nn\\
Y^{t_{n}}_I &=&  -Y^e_I \frac{\partial^2 S_{n} }{\partial l^n_e \partial t^{n}}\left(   \frac{\partial^2 S_{n} }{\partial t_{n} \partial  t_{n}}     \right)^{-1}    \q .
\ea
(Here we assume that the second partial derivatives of 
the action with respect to $t_{n-1}$ and $t_n$ do not vanish. This is generically the case for the Regge 
action.)

The first equation in (\ref{oct9}) together with (\ref{oct7a}) and (\ref{oct9a}) can be used to show that the components $\{Y^e_I\}$ constitute null vectors for the Hessian of the effective action $\tilde S$, that is
\ba\label{oct10}
%\sum_e
&& Y^e_I
\frac{\partial^2 \tilde S }{\partial l_n^e \partial l^{e'}_n} \nn\\
&=& 
 Y^e_I
 \left( 
 \frac{\partial^2  S }{\partial l_n^e \partial l^{e'}_n} 
 - \frac{\partial^2  S_{n-1} }{\partial l_n^e \partial t_{n-1}} \left(  \frac{\partial^2  S_{n-1} }{\partial t_{n-1} \partial t_{n-1}} \right)^{-1} 
\frac{\partial^2  S_{n-1} }{\partial t_{n-1} \partial l^{e'}_n}
-\frac{\partial^2  S_{n} }{\partial l_n^e \partial t_n} \left(  \frac{\partial^2  S_{n} }{\partial t_n \partial t_n} \right)^{-1} 
\frac{\partial^2  S_{n} }{\partial t_n \partial l^{e'}_n}
\right)  \nn\\ &=&  0 \q . 
\ea

Later on, in section \ref{dyn}, we will need this relation in the form
\ba\label{oct10a}
Y^e_I  \frac{\partial^2 \tilde S_{n-1} }{\partial l_n^e \partial l^{e'}_n} + Y^e_I  \frac{\partial^2 \tilde S_{n} }{\partial l_n^e \partial l^{e'}_n} =0 \q .
\ea
%where the second derivative of $\tilde S_n$ is explicitly given by
%\ba\label{oct10b}
%\frac{\partial^2 \tilde S_{n} }{\partial l_n^e \partial l^{e'}_n} =\frac{\partial^2  S_{n} }{\partial l_n^e \partial l^{e'}_n} - 
%\frac{\partial^2  S_{n} }{\partial l_n^e \partial t_n} \left(  \frac{\partial^2  S_{n} }{\partial t_n \partial t_n} \right)^{-1} 
%\frac{\partial^2  S_{n} }{\partial t_n \partial l^{e'}_n}
%\ea
%and similarly for $\tilde S_{n-1}$. Note that the $Y^e_I$ 
Similarly, we have 
\ba\label{oct10c}
Y^e_I  \frac{\partial^2 \tilde S_{n-1} }{\partial l_n^e \partial l^{b}} + Y^e_I  \frac{\partial^2 \tilde S_{n} }{\partial l_n^e \partial l^{b}} =0  \q,
\ea
where $l^b$ is the length of an edge which is contained in both three-dimensional $\text{star}(v_{n-1})$ and $\text{star}(v_{n})$.  This equation follows if we consider the Hessian associated to a larger boundary problem, also including $l^b$ as free variable. Since a translation of a vertex $v_n$ only affects the lengths of the edges adjacent to $v_n$ this Hessian still has the null vectors $Y_I^\iota$ (with the components $\iota \neq e,t_{n-1},t_n$ vanishing), in particular,
\ba\label{oct10d}
 Y^e_I \frac{\partial^2 S }{\partial l^e_n
\partial l^{b}}
 + Y^{t_{n-1}}_I  \frac{\partial^2 S }{\partial t_{n-1} \partial l^{b}}
+ Y^{t_{n}}_I  \frac{\partial^2 S }{\partial t_{n} \partial l^{b}}  
&=&0  \q .
\ea
Together with the equations (\ref{oct9a}) for the components $Y^{t_{n-1}}_I$ and $Y^{t_{n}}_I$, we obtain the result asserted in equation (\ref{oct10c}).

%The equations in (\ref{oct9a}) express the components $Y^{t_
%{n-1}}$ 
%and $Y^{t_n}$
%as a combination of the components $Y^ e$ . For this reason a set of linearly independent null vectors for the Hessian of $S$ will define
%a set of independent null vectors for the Hessian of $\tilde S$ of the same size. 

Next, we will show that the null vectors $Y^e_I$ are also left or right null vectors of 
\ba\label{need1}
\frac{\partial^2 \tilde S_n}{\partial l^e_n \partial l^{e'}_{n+1}} \q \text{or} \q  \frac{\partial^2 \tilde S_{n-1}}{\partial l^e_{n-1} \partial l^{e'}_{n}}\q,
\ea
respectively. %Here $\tilde S_n$ is the action $S_n$ with the variable $t_n$ replaced by the solution $T_n(l^e_{n},l^e_{n+1})$. 
 Assume that we extremize the action $\tilde S$ with respect to the variables $l^e_n$, fixing the variables $l^e_{n-1}$ and $l^e_{n+1}$. Calling the corresponding solutions $L^e_n(l^{e}_{n-1},l^{e'}_{n+1})$, we obtain the identities
\ba\label{oct11}
\frac{\partial \tilde S}{\partial l_n^e}(l^e_{n-1},L_n^e( l^e_{n-1},l_{n+1}^e  ), l^e_{n+1}) \equiv 0  \q .
\ea 
Differentiating these equations with respect to $l^{e'}_{n+1}$ or $l^{e'}_{n-1}$ results in
\ba
\frac{\partial^2 \tilde S}{\partial l_n^e \partial l_{n}^{e''} } \frac{\partial L^{e''}_{n}}{\partial l^{e'}_{n+1}}
+ \frac{\partial^2 \tilde S_n}{\partial l_n^e \partial l_{n+1}^{e'} } &=&0\; ,    \nn\\
 \frac{\partial L^{e''}_{n}}{\partial l^{e'}_{n-1}} \frac{\partial^2 \tilde S}{\partial l_n^{e''} \partial l_{n}^{e} }
+ \frac{\partial^2 \tilde S_{n-1}}{\partial l_{n-1}^{e'} \partial l_{n}^{e} } &=&0 \q,
\ea
respectively. This proves the claim.

Finally, let us note that the components $Y^e_I$ can be computed from the lengths of the edges in the 3d  $\text{star}(v_n)$ only. As was noted in section \ref{bianchi}, we have $Y^e_I=\vec{B}_I\cdot \vec{E}^e/|\vec{E^e}_I|$ where $B_I$ is a basis in 4d flat space and $\vec{E}^e$ are the 4d vectors for the edges $e$, which one can obtain by embedding $\text{star}(v_n)$ into 4d flat space. From a counting of variables argument, which can be found in section \ref{sechigher}, one can deduce that the lengths in  $\text{star}(v_n)$ determine this embedding uniquely (modulo translations and rotations). Hence, the components $Y^e_I$ are determined by $l^e_n$ and $l^b$. As we will see in the next sections, the same holds for the constraints of the linearized theory, i.e.\ these depend only on the (fluctuation and background) variables associated to a particular (discrete) time.

\section{The linearized theory}\label{lin}

Here we have to consider the tent move equations (\ref{oct3}) to linear order. That is, we expand all lengths $l^e$ of the triangulation around a flat solution ${}^{(0)}l^e$ in a small parameter $\varepsilon$ and keep only terms linear in $\varepsilon$ in the equations. Later on, we will also discuss an expansion to higher order in $\varepsilon$.

Among the possible variations of the lengths are also the ones in flat directions. These variations are solutions of the linearized equations of motion. As solutions to linear equations of motions are additive these variations can also be added to solutions representing linearized curvature excitations without changing the boundary data. All the solutions of the linearized theory will therefore exhibit gauge symmetries. In general, the linearized theory will inherit the gauge freedom of the background solution.

In a tent move we encounter three different kind of edges. Firstly, there are edges $b$ in the intersection of the two Cauchy surfaces $\Sigma_n \cap \Sigma_{n+1}$ defined by the tent move. 
These edges are not dynamical (for this specific tent move) and we will call the associated length variables $l^b={}^{(0)}l^b+ \varepsilon y^b+ O(\varepsilon^2)$. 
Secondly, there are the 'dynamical edges' in the Cauchy surfaces $\Sigma_n$ and $\Sigma_{n+1}$ adjacent to the evolving vertices $v_n,v_{n+1}$. The associated length variables will be $l^e={}^{(0)}l^e+ \varepsilon y^e+ O(\varepsilon^2)$.  Finally, there is the tent pole -- the only bulk edge -- with length $t_n={}^{(0)}t^n+ \varepsilon x_n+O(\varepsilon^2)$.

Furthermore, we expand the momenta conjugate to $l^e_n$ and $t_n$, that is $p_e^n={}^{(0)}p_e^n + \varepsilon \pi_e^n+ O(\varepsilon^2)$ and $p_t^n={}^{(0)}p_t^n + \varepsilon \pi_t^n+ O(\varepsilon^2)$. To linear order in $\varepsilon$ we obtain from the first two equations in (\ref{oct3}), which define the momenta $p^n_e$ and $p^n_t$,
\ba\label{lin2}
\pi^n_e      %&:=& -\frac{\partial\,\,\, {}^{[2]} \!S_{(n,n+1)}}{\partial\, \varepsilon y^n_e} \nn\\
&=&        % -\frac{\partial S_{(n,n+1)}} {\partial l^n_e}
-  \frac{\partial^2 S_n}{\partial l^e_n\partial l^{e'}_n} y_n^{e'}
 -   \frac{\partial^2 S_n}{\partial l^e_n\partial l^{e'}_{n+1}} y_{n+1}^{e'}
-  \frac{\partial^2 S_n}{\partial l^e_n\partial t_n} x_n
- \frac{\partial^2 S_n}{\partial l^e_n\partial l^b} y^b \q ,\\
\pi^n_t   %  &:=& -\frac{\partial \,\,\, {}^{[2]} \!S_{(n,n+1)}}{\partial\, \varepsilon x_n}   \label{lin3} \nn\\
&=& 
- \frac{\partial^2 S_n}{\partial t_n\partial l_n^{e}} y_{n}^e
-  \frac{\partial^2 S_n}{\partial t_n\partial l^{e}_{n+1}} y_{n+1}^{e}
-\varepsilon\frac{\partial^2 S_n}{\partial t_n\partial t_n} x_n  
-\frac{\partial^2 S_n}{\partial t_n\partial l^b} y^b       \label{lin3} \nn\\
&=&0\q ,
\ea
where it is understood that the derivatives of the action $S_n$ are evaluated on the flat background solution.

Solving equation (\ref{lin3}) for the tent pole variable $x_n$ and using this in (\ref{lin2}), we obtain
\ba\label{lin4}
\pi^n_e &=&     %-\frac{\partial S_{(n,n+1)}} {\partial l_{e_n}} \nn\\
%&&
%-\varepsilon \sum_{e'} \left( \frac{\partial^2 S_{(n,n+1)}}{\partial l_e^n\partial l_{e'}^{n+1}} 
%- \frac{\partial^2 S_{(n,n+1)}} {\partial l_{e}^n\partial t^n} \left( \frac{\partial^2 S_{(n,n+1)}}{\partial t^n\partial t^n}    \right)^{-1}
%\frac{\partial^2 S_{(n,n+1)}}{\partial t^n\partial l^{n+1}_{e'}}\right) y^{n+1}_{e'}   \nn\\&&
-M^n_{ee'} y_{n+1}^{e'}
%&&
%
%- \left( \frac{\partial^2 S_{(n,n+1)}}{\partial l_e^n\partial l_{e'}^n} 
%- \frac{\partial^2 S_{(n,n+1)}} {\partial l_{e}^n\partial t^n} \left( \frac{\partial^2 S_{(n,n+1)}}{\partial t^n\partial t^n}    \right)^{-1}
%\frac{\partial^2 S_{(n,n+1)}}{\partial t^n\partial l^n_{e'}}\right) y^{n}_{e'}   \nn\\
-N_{ee'}^n y_n^{e'} 
%-\varepsilon \sum_{b} \left( \frac{\partial^2 S_{(n,n+1)}}{\partial l_e^n\partial l_b} 
%- \frac{\partial^2 S_{(n,n+1)}} {\partial l_{e}^n\partial t^n} \left( \frac{\partial^2 S_{(n,n+1)}}{\partial t^n\partial t^n}    \right)^{-1}
%\frac{\partial^2 S_{(n,n+1)}}{\partial t^n\partial l_b}\right) r_b 
-N_{eb}^n y^b\q ,
%\q .\q\q
\ea
where
\ba\label{lin5}
M^n_{ee'}&=& \frac{\partial^2 S_{n}}{\partial l^e_n\partial l^{e'}_{n+1}} 
- \frac{\partial^2 S_{n}} {\partial l_n^{e}\partial t_n} \left( \frac{\partial^2 S_{n}}{\partial t_n\partial t_n}    \right)^{-1}
  \frac{\partial^2 S_{n}}{\partial t_n\partial l_{n+1}^{e'}}  \q =  \q\frac{\partial^2 \tilde S_{n}}{\partial l^e_n\partial l^{e'}_{n+1}}  \nn\\
  N^n_{ee'}&=& \frac{\partial^2 S_{n}}{\partial l^e_n\partial l^{e'}_{n}} 
- \frac{\partial^2 S_{n}} {\partial l_n^{e}\partial t_n} \left( \frac{\partial^2 S_{n}}{\partial t_n\partial t_n}    \right)^{-1}
  \frac{\partial^2 S_{n}}{\partial t_n\partial l_{n}^{e'}} \q\;\,\,\q= \q \frac{\partial^2 \tilde S_{n}}{\partial l^e_n\partial l^{e'}_{n}}  \nn\\
  N^n_{eb}&=& \frac{\partial^2 S_{n}}{\partial l^e_n\partial l^{b} }
- \frac{\partial^2 S_{n}} {\partial l_n^{e}\partial t_n} \left( \frac{\partial^2 S_{n}}{\partial t_n\partial t_n}    \right)^{-1}
  \frac{\partial^2 S_{n}}{\partial t_n\partial l_{}^{b}} \;\;\;\,\;\q\q= \q\frac{\partial^2 \tilde S_{n}}{\partial l^e_n\partial l^{b} } \q .
\ea

From the discussion in section \ref{deg} we know that the first matrix $M^n_{ee'}$ has four (left) null eigenvectors $Y_I^e,I=1,\ldots,4$. Contracting equation (\ref{lin4}) with the left null eigenvectors we find four relations involving only momenta and configuration variables at time step $n$, that is constraints
\ba\label{lin6}
C_I&=&
Y^e_I \pi_e^n+
%\sum_e Y^e_I\pi^n_e  +
%\sum_e Y^e_I \frac{\partial S_{(n,n+1)}} {\partial l_{e_n}} \nn\\
Y^e_I
%\left( \frac{\partial^2 S_{(n,n+1)}}{\partial l_e^n\partial l_{e'}^n} 
%- \frac{\partial^2 S_{(n,n+1)}} {\partial l_{e}^n\partial t^n} \left( \frac{\partial^2 S_{(n,n+1)}}{\partial t^n\partial t^n}    \right)^{-1}
%\frac{\partial^2 S_{(n,n+1)}}{\partial t^n\partial l^n_{e'}}\right) 
\,N^n_{ee'}\,
y_{n}^{e'}   
+Y^e_I
%\left( \frac{\partial^2 S_{(n,n+1)}}{\partial l_e^n\partial l_b} 
%- \frac{\partial^2 S_{(n,n+1)}} {\partial l_{e}^n\partial t^n} \left( \frac{\partial^2 S_{(n,n+1)}}{\partial t^n\partial t^n}    \right)^{-1}
%\frac{\partial^2 S_{(n,n+1)}}{\partial t^n\partial l_b}\right) 
\,N^n_{eb} \,
y^b   \,\, =\,\,0 \q .
\ea
These constraints have to be satisfied by the canonical data of the linearized theory in order to give rise to a solutions of the equations of motion. Although the constraints involve only the momenta and configuration variables of linear order, variables describing the flat background solution might appear, which also involve data at the time step $(n+1)$. However, these could be replaced--using the equation of motion for the background--by the configuration variables and momenta at time step $n$ as well as background lapse and shift (describing the gauge freedom of the background solution). But, as we will see, this issue will not arise. That is, the constraints will only involve the background variables at time step $n$ and not depend on the position of the vertex $v_{n+1}$ of the flat background solution.

%Note that also the momenta $\pi^n_e$ could be expanded as a series in $\epsilon$. Then the constraints (\ref{lin6}) contain the zeroth order constraints and the first order constraints. If we use the (canonical) equations of motion for the background variables then the zeroth order constraints are automatically satisfied. 
%Also by using these equations we can rewrite any variables ${}^fl^{n+1}$ that might appear in the derivatives of the action (evaluated on a flat background solution) as a function of ${}^fl^n$ and ${}^f\pi^n$ and the lapse and shift parameters. (These lapse and shift parameters should actually drop out in the end.) In this way one can obtain the constraints for an expansion around a general flat background. {\it One has to be a bit careful about using equations for the flat background as that might redefine what one means with zeroth order constraints.}

%{\bf Remark:} We could also introduce the `effective action' $\tilde S_{(n,n+1)}$ which is $S_{(n,n+1)}$ with the $t_n$ variable integrated out and rewrite everything using this action, see section 7 in paper 1 and see below. Then all these terms in big brackets can be replaced by double derivatives of the action.

\section{The constraints at a four--valent vertex}  \label{4valent}

\begin{figure}[hbt!]
\begin{center}
    \psfrag{v0}{$v_0$}
    \psfrag{v1}{$v_1$}
    \psfrag{1}{$1$}
    \psfrag{2}{$2$}
    \psfrag{3}{$3$}
    \psfrag{lb}{$l_b$}
    \includegraphics[scale=0.4]{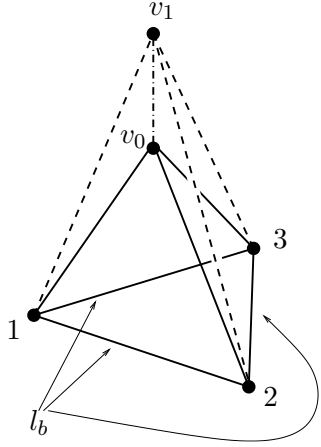}
    \end{center}
    \caption{\label{3dan}{\small Construction of the 3d three-valent analogue of the 4d four-valent tent move where both tetrahedra are oriented in the same (future) direction.}}
\end{figure}

Here we will discuss the constraints (\ref{lin6}) in more detail and derive a more explicit expression. We will start with constraints derived from the tent move at a four--valent vertex. A four--valent vertex in the 3d boundary of a 4d triangulation can be identified as  a vertex of a 4--simplex of this 4d triangulation. For a four--valent tent move one can\footnote{Under the condition that the triangle inequalities are satisfied in the construction below .} construct a flat solution.\footnote{There might also exist exceptional cases where solutions with curvature are possible, see the discussion in \cite{bahrdittrich09a}. These seem to be discretization artifacts, however, and we will ignore these kind of solutions. Moreover, also the flat solutions might be ambiguous, in particular there will be future directed and  past directed solutions. We will always choose the future directed solution.} Indeed, having given the four edge lengths $l_e^n$ and $l_e^{n+1}, e=e(v1),\ldots,e(v4)$ in addition to the six edge lengths $l_b, b=e(12),\ldots,e(34)$ we can construct a solution by taking two 4--simplices $\sigma(v_n1234)$ and $\sigma(v_{n+1} 1234)$ with the appropriate edge lengths and gluing these together along $\tau(1234)$. Connecting $v_n$ with $v_{n+1}$, we will obtain the length of the tent pole $t_n$. All the deficit angles at the triangles hinging at the tent pole vanish, hence the Regge equation associated to the tent pole is satisfied.  The analogous construction in 3d is depicted in figure \ref{3dan}. 

\noindent

With the solution  $T_n(l_n^e,l_{n+1}^e,l^b)$ of the tent pole equation,
\ba\label{sep5}
0=p^t_n=-\frac{\partial S_{n}}{\partial t_n}= \sum_{\Delta \supset t_n} \frac{\partial A_\Delta}{\partial t_n} \epsilon_{\Delta} \q ,
\ea
 we can define the following functions of $l^{e}_n,l^{e}_{n+1}$ and $l^b$
\ba\label{sep4}
\tilde p_{e}^n&:=&{p_e^n}_{|t=T_n(l_n^e,l_{n+1}^e,l^b)}:=-{\frac{\partial S_{n}}{\partial l^e_n}}_{|t=T_n(l_n^e,l_{n+1}^e,l^b)}\nn\\
\tilde p^n_b &:=&
 {p^n_b}_{|t=T_n(l_n^e,l_{n+1}^e,l^b)}:=
-{\frac{\partial S_{n}   }{\partial l_b}}_{|t=T_n(l_n^e,l_{n+1}^e,l^b)} \q .
\ea
%where $t_s(l^n_e,l^{n+1}_e,l_b)$ denotes the solution to the equation for the `tent pole' edge
Taking the derivative of equation (\ref{sep5}) with respect to $l_e^n$, we obtain
\be\label{sep6}
\frac{\partial T_n}{\partial l^e_n}=-\left(\frac{\partial^2 S_{n}}{\partial t_n\partial t_n} \right)^{-1} \frac{\partial^2 S_{n}}{\partial t_n l^e_n}
\ee
and can conclude that the matrices $N^n_{e'e}$, $N^n_{e'b}$ defined in (\ref{lin5}) can be expressed as
\ba\label{sep7}
\frac{\partial \tilde p_e^n}{\partial l^{e'}_n}= -N^n_{e'e}\q , \q\q
\frac{\partial \tilde p_b^n}{\partial l^{e'}_n}= -N^n_{e'b}   \q .
\ea

As discussed above, for the four--valent tent move the deficit  angles at the triangles hinging at the tent pole vanish, hence the expressions for $\tilde p^e_n,\tilde p^b$ simplify to
\ba\label{sep9}
{\tilde p^e_n}&=&\sum_{\Delta \subset  \overset{\circ}{\Sigma}_n} \frac{\partial A_\Delta}{\partial l^e_n} \,\psi_\Delta(l_n^{e'},l_{n+1}^{e'},l_b)  \nn\\
{\tilde p^b}&=& \sum_{\Delta \subset \Sigma_n \cup \Sigma_{n+1}}   \frac{\partial A_\Delta}{\partial l^b} \,      \psi_\Delta(l_n^{e'},l_{n+1}^{e'},l^b)   \q .
\ea

Here $\psi_\Delta(l^n_{e'},l^{n+1}_{e'},l_b)$ are the extrinsic curvature angles for the boundary triangles of a piece of flat triangulation.  For the situation depicted in figure \ref{3dan} these angles are given by 
\begin{alignat}{2}\label{sep10}
\psi_\Delta &= -\pi+\theta_\Delta(l_n^{e'},l^b)\q\q &&\text{for} \, \Delta \subset \overset{\circ}{\Sigma}_n\nn \\
\psi_\Delta &=    \pi -\theta_\Delta(l_{n+1}^{e'},l^b)\q\q&&\text{for} \, \Delta \subset \overset{\circ}{\Sigma}_{n+1}\nn \\
\psi_\Delta &= -\theta_\Delta(l_{n+1}^{e'},l^b)+\theta_\Delta(l_n^{e'},l^b)\q\q
&&\text{for}  \, \Delta \subset {\Sigma_n \cap \Sigma_{n+1}}\q ,
\end{alignat}
where $\theta_\Delta(l)$ is the dihedral angle on a 4--simplex $\sigma$ with edge lengths $l$. 

In order to compute the  matrices $N_{e'e}^n$ and $N_{e'b}^n$, note that taking the derivative with respect to $l^{e'}_n$ will annihilate the dihedral angles $\theta_\Delta(l_{n+1}^{e'},l^b)$ in (\ref{sun1}), so that everything can be expressed with dihedral angles of the simplex $\sigma$ with edge lengths $(l_n^e,l^b)$:
\ba\label{sun1}
N_{e'e}^n&=&\frac{\partial }{\partial l^{e'}_n} \sum_{\Delta \subset  \sigma} \frac{\partial A_\Delta}{\partial l^e_n} \, ( \pi-  \theta_\Delta(l_n^{e'},l^b))\nn\\
 & =&  \frac{\partial }{\partial l^{e}_n} \sum_{\Delta \subset  \sigma} \frac{\partial A_\Delta}{\partial l^{e'}_n} \, ( \pi-\theta_\Delta(l_n^{e'},l^b))                   \, ,
 \ea
 where we used the fact that the right--hand sides of both sides is a double derivative of $S_\sigma=\sum_\Delta A_\Delta(\pi-\theta_\Delta)$. Formulae for the derivatives of dihedral angles can be found in \cite{dittrichfreidelspeziale}.
 
 The same applies to the computation of $N_{e'b}^n$, moreover, note that $\partial/\partial l^{e'}_n A_{\Delta(ijk)}=0$ as $A_{\Delta(ijk)}$ only depends on the lengths $l^b$. Thus, we may write
 \ba\label{sun2}
N_{e'b}^n&=&\frac{\partial }{\partial l^{e'}_n} \left( \sum_{\Delta \subset  \sigma, \Delta=\Delta(v_0ij)} \frac{\partial A_\Delta}{\partial l^b} ( \pi-\theta_\Delta(l_n^{e'},l^b))\,\,\,\,  - \sum_{\Delta \subset  \sigma, \Delta=\Delta(ijk)} \frac{\partial A_\Delta}{\partial l^b} \theta_\Delta(l_n^{e'},l^b) \right)\nn\\
&=& \frac{\partial }{\partial l^{b}} \sum_{\Delta \subset  \sigma} \frac{\partial A_\Delta}{\partial l^{e'}_n} \, ( \pi-\theta_\Delta(l_n^{e'},l^b))  \q .
\ea
Note that the dependence of the matrices $N_{e'e}^n,N^{e'b}_n$  on the lengths $l^e_{n+1}$ drops out completely. That is, we can compute the constraints from the (background) geometrical data on $\Sigma_n$ only.

To finally compute the constraints 
\ba
C_I=Y_I^e (\pi^n_e+ N_{ee'}^n y^{e'}+N_{eb}y^b)
\ea
 we need to determine the four vector fields $Y_I^e$. As we have four vector fields and four edges $e=e(v_0i)$ we can identify the indices $I$ and $e$ and define $Y_{e'}^e$ to be the vector field that translates the vertex $v_0$ orthogonal to the other three edges $e^{''}\neq e'$.  It is easy to see that (assuming normalization) we have $Y_{e'}^e=\delta^e_{e'}$. This determines the first order constraints for a four-valent vertex to
\ba\label{4vconstraints}
C_e=\pi_e^n +  \frac{\partial }{\partial l^{e}_n} \sum_{\Delta \subset  \sigma} \frac{\partial A_\Delta}{\partial l^{e'}_n} \, ( \pi-\theta_\Delta(l_n^{e'},l^b))      y^e_n +    \frac{\partial }{\partial l^{b}} \sum_{\Delta \subset  \sigma} \frac{\partial A_\Delta}{\partial l^{e'}_n} \, ( \pi-\theta_\Delta(l_n^{e'},l^b))y^b \q .
\ea

Note that these constraints coincide with the first order of the full constraints derived in \cite{dittrichryan, dittrich08}
\ba\label{flata}
C_e^{full}=p_e^n+ \sum_{\Delta\subset \sigma} \frac{\partial A_\Delta}{\partial l^e_n} (\pi- \theta_\Delta(l^{e'}_n,l^b)) \q .
\ea 
These constraints for the four valent vertex can be generalized to the `flat sector' of 4d Regge calculus, which are a special class of triangulated hypersurfaces that only allow for evolution leading to flat 4d space, see \cite{dittrichryan}.

\section{The constraints at higher valent vertices}\label{sechigher}

Next, we will derive the constraints at higher valent vertices. The discussion will be in many aspects parallel to the one in the last section \ref{4valent}. There is one important difference however, which is that solving the tent pole equation for higher valent vertices can also lead to solutions with non--vanishing deficit angles. Nevertheless, we will see that again for the computation of the constraints we will only need the geometrical data of the hypersurface $\Sigma_n$.

To this end, note that a tent move at a vertex $v$ only involves the 3d star of this vertex, i.e.\ all simplices (with their subsimplices)  that share the vertex $v$. As we perform an expansion around flat space the configuration we are considering must be embeddable into flat 4d space.  Indeed, there are as many edge lengths in the 3d star of a vertex as one needs to determine an embedding into flat 4d space modulo translations and rotations  \cite{rocekwilliams}. Firstly, we will count the number of edges in the star of an $N$--valent vertex. In addition to the $N$ edges adjacent to $v$ we have $E$ edges in the boundary of the star. The piecewise linear manifold condition \cite{ambjornbook} ensures that this boundary is topologically a  2--sphere. For the number of edges $E$, the number of triangles $T$ and the number of vertices $V$ in a triangulated 2--sphere there are two relations, the Euler theorem $T-E+V=2$ and the relation $3T=2E$. Hence, the number of edges in the 2--sphere is $E=3V-6$. The number of vertices is $V=N$ so that the overall number of edges in the 3d star is $4N-6$. On the other hand, if we embed  the 3d star of the $N$--valent vertex $v$ into 4d flat space we have to choose $4N$ coordinates for the $N$ vertices. Modulo the six 4d rotations this will also give $4N-6$ parameters. Hence, one can expect that the lengths of the edges in the 3d star uniquely determine an embedding into flat 4d space. (There might occur discrete ambiguities, however, these are fixed by the flat background solution under consideration.) 

To derive the constraints we will again use the momenta $\tilde p^n_e,\tilde p^n_b$ introduced in (\ref{sep4}) in order to compute the matrices $N^n_{e'e}$ and $N^n_{e'b}$. These momenta involve the solution of the tent pole equation (\ref{sep5}) which, however, can now lead to solutions with non--vanishing deficit angles. That is, we have
\ba\label{higher1}
{\tilde p^e_n}{}_{|flat}&=&\sum_{\Delta \subset  \overset{\circ}{\Sigma}_n} \frac{\partial A_\Delta}{\partial l^e_n} \,\psi_\Delta(l_n^{e'},l_{n+1}^{e'},l^b)  \nn\\
{\tilde p^b}{}_{|flat}&=& \sum_{\Delta \subset \Sigma_n \cup \Sigma_{n+1}}   \frac{\partial A_\Delta}{\partial l^b} \,      \psi_\Delta(l_n^{e'},l_{n+1}^{e'},l^b)   \q ,
\ea
only on data $(l^e_n,l^e_{n+1},l^b)$ which lead to a solution of the tent pole equation with vanishing deficit angles. (Fixing $(l^e_n,l^b)$, that is the geometry of the 3d star $\Sigma_n$, there is generically a 4--parameter set of lengths $l^e_{n+1}$ such that one can obtain a flat solution.) 

This is, however, sufficient to compute the contraction of the matrices $N^n_{e'e}, N^n_{e'b}$ with the vectors $Y_I^e$, which is all we need to determine the constraints. According to (\ref{sep7}), this contraction corresponds to the derivatives of the momenta $\tilde p$ in the direction of $Y_I^e$:
\ba\label{higher2}
Y_I^{e'}\frac{\partial \tilde p_e^n}{\partial l^{e'}_n}= -Y^{e'}_IN^n_{e'e}\q , \q\q
Y_I^{e'}\frac{\partial \tilde p_b^n}{\partial l^{e'}_n}=-Y_I^{e'} N^n_{e'b}   \q .
\ea
As explained in section \ref{bianchi}, these vectors correspond to translations of the vertex $v_n$ and the induced change of lengths $l^e_n$ such that the triangulation remains flat. Hence, we can still use the expression (\ref{higher1}) to determine the derivatives in `flat directions'. As for the 4--valent vertex we have again
\begin{alignat}{2}\label{higher3}
\psi_\Delta &= -\pi+\theta_\Delta(l_n^{e},l^b)\q\q &&\text{for} \, \Delta \subset \overset{\circ}{\Sigma}_n\nn \\
\psi_\Delta &=    \pi -\theta_\Delta(l_{n+1}^{e},l^b)\q\q&&\text{for} \, \Delta \subset \overset{\circ}{\Sigma}_{n+1}\nn \\
\psi_\Delta &= -\theta_\Delta(l_{n+1}^{e},l^b)+\theta_\Delta(l_n^{e},l^b)\q\q
&&\text{for}  \, \Delta \subset {\Sigma_n \cap \Sigma_{n+1}}\q ,
\end{alignat}
where we now generalized the angles $\theta_\Delta$ to the dihedral angle between the two tetrahedra sharing the triangle $\Delta$ of the 3d star of $v_n$ or $v_{n+1}$, respectively, embedded into 4d flat space. As discussed above, these embeddings, and hence the dihedral angles, are determined by the edge lengths $(l^e_n,l^b)$  or $(l^e_{n+1},l^b)$, respectively. Again, the dihedral angles $\theta_\Delta(l_{n+1}^{e},l^b)$ drop out after taking the derivatives with respect to $l^{e'}_n$. As explained in section \ref{deg}, the vectors $Y_{I}^e$ can be determined as functions of $l^{e}_n, l^b$. The linearized constraints take the form  
\ba\label{higher4}
C_I &=&Y_{I}^{e'} \pi^n_{e'} + Y^{e'}_I  \frac{\partial}{\partial l^{e'}_n} \sum_{\Delta \subset  \overset{\circ}{\Sigma}_n}\frac{\partial A_\Delta}{\partial l^e_n} \, (\pi - \theta_{\Delta} (l^{e}_n,l^b)) y^{e}_n    \nn\\
&&\q\q\;\,\,+Y^{e'}_I  \frac{\partial}{\partial l^{e'}_n} \sum_{\Delta \subset  {\Sigma}_n}\frac{\partial A_\Delta}{\partial l^b} \, (\pi-\theta_{\Delta} (l^{e}_n,l^b) )y^b \, . \q\q
\ea

Also here the linearized constraints do not depend on the lengths in the background solution at the next time step, in particular they do not depend on background lapse and shift.  In this sense the linearized constraints are independent from the tent move construction: they refer only to the data on the Cauchy surface in question, hence we do not need to specify any tent moves in order to state the constraints.

The constraints  (\ref{higher4}) also generate the expected gauge transformations. The constraint $C_I$ is expected to generate the change of coordinates induced by translating the vertex $v_n$ in the direction determined by $Y_I^e$ (whose components describe the induced change of length variables). Indeed, 
\ba\label{higher5}
\{y^e_n, C_I\}&=&Y^{e}_I  \q .
\ea
To determine the infinitessimal change of the momenta, remember that translating a vertex in the embedding flat space does not change the flatness of the configurations. On such flat configurations the momenta are given by (\ref{higher1}). (For the boundary edges with index $b$ formula (\ref{higher1}) gives the part that does depend on the length $l^e_n$ of the edges adjacent to $v_n$.) Therefore, 
\ba\label{higher6}
\{\pi^e_n,C_I\} &=&- Y^{e'}_I  \frac{\partial}{\partial l^{e'}_n} \sum_{\Delta \subset  \overset{\circ}{\Sigma}_n}\frac{\partial A_\Delta}{\partial l^e_n} \, (\pi - \theta_{\Delta} (l^{e}_n,l^b))  \nn\\
\{\pi^b,C_I\}&=&-Y^{e'}_I  \frac{\partial}{\partial l^{e'}_n} \sum_{\Delta \subset  {\Sigma}_n}\frac{\partial A_\Delta}{\partial l^b} \, (\pi-\theta_{\Delta} (l^{e}_n,l^b))
\ea
reproduces the correct transformation behaviour for the linearized momenta. Hence, requiring that the constraints $C_I$ generate the change of variables induced by vertex translations in the direction $Y_I^e$ gives an alternative derivation of the formula (\ref{higher4}).

\subsection{Example: five--valent symmetry--reduced vertex}\label{beispiel1}

As an example we consider a tent move at a five--valent vertex. To simplify the situation, we will consider a `symmetry--reduced' set--up, also used in \cite{bahrdittrich09a}, so that we only have two dynamical length variables, $a_n$ and $b_n$, to deal with at each time step.

\begin{figure}[hbt!]
\begin{center}
    \psfrag{v}{$v$}
    \psfrag{a}{$a$}
    \psfrag{b}{$b$}
    \psfrag{1}{$1$}
    \psfrag{2}{$2$}
    \psfrag{3}{$3$}
    \psfrag{4}{$4$}
    \psfrag{5}{$5$}
    \includegraphics[scale=0.4]{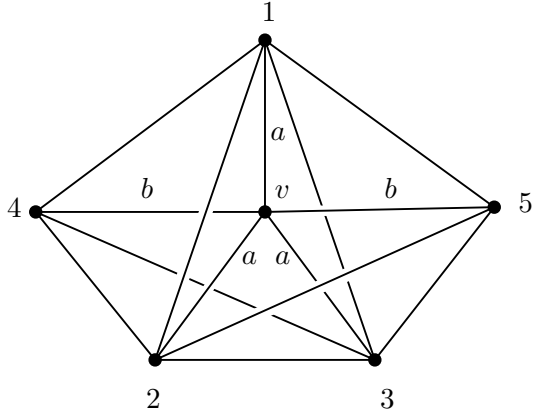}
    \end{center}
    \caption{\label{5vsym}{\small Illustration of the symmetry--reduced 3d $\text{star}(v)$ of a five--valent vertex $v$, consisting of the six tetrahedra $ \tau(v124), \ \tau(v134), \ \tau(v234), \ \tau(v125), \ \tau(v135)$ and 
$\tau(v235)$.}}
\end{figure}

The geometry of the 3d $\text{star}(v)$ is illustrated in figure \ref{5vsym} and given as follows:\\
 As $v$
is five--valent we have five further vertices which we will denote
by $1,\ldots,5$. We will assume that we have six tetrahedra with
vertices \be\label{tent1} v124,\q v134,\q v234, \q  v125, \q  v135,
\q v235  \q . \ee Accordingly, we will have nine triangles of the
form $\Delta(vij)$ with $i,j=1,\ldots 5$ in this triangulation, five edges
of the form $e(vi)$ and nine edges of the form $e(ij)$  (all possible
ordered combinations of $i,j\in \{1,\ldots 5\}$ with the exception
$45$).

\noindent

The symmetry assumption includes to set all the lengths of the boundary edges $e(ij)$ to $1$ \footnote{This is possible since the vacuum Regge equations are invariant under global rescalings.} (and we will neglect variations $y^b$ at these boundary edges) and setting equal $l^{e(vi)}=a$, $i=1,2,3$  as well as $l^{e(v4)}=l^{e(v5)}=b$.

The 4--simplices involved in the tent move are then all of the same type
$\sigma(v_0v_1ij\kappa)$ where $v_0,v_1$ denote the two vertices of the tent
pole (at time steps $n=0,n=1$, respectively), $i,j$ take values in
$1,2,3$ and $\kappa$ in $4,5$. We will denote by
\begin{itemize} \parskip -3pt
\item[~] $\theta^0_a,\, A^0_a$ the dihedral angle and the area of the triangle $\Delta(v_0ij)$,
\item[~] $\theta^0_b,\, A^0_b$ the dihedral angle and the area of the triangle $\Delta(v_0i\kappa) $,
\item[~] $\theta^a_t,\, A^a_t$ the dihedral angle and the area of the triangle $\Delta(v_0v_1i )$,
\item[~] $\theta^b_t,\, A^b_t$ the dihedral angle and the area of the triangle $\Delta(v_0v_1\kappa) $,
\item[~] $\theta^1_a,\, A^1_a$ the dihedral angle and the area of the triangle $\Delta(v_1ij)$,
\item[~] $\theta^1_b,\, A^1_b$ the dihedral angle and the area of the triangle $\Delta(v_1i\kappa)$ 
\item[~] $\theta,\, A$ the dihedral angle and the area of the triangle $\Delta(ij\kappa)$, respectively .
\end{itemize}

The Regge action for one time step is given by
\ba\label{novemb1}
S_0&=&-3A_t^a(2\pi- 4 \theta^a_t) - 2 A^b_t(2\pi-3\theta^b_t) - 3 A^0_a(\pi-2\theta^0_a)-6 A^0_b(\pi-2 \theta^0_b)\nn\\
&&- 3 A^1_a(\pi-2\theta^0_a)-6 A^1_b(\pi-2 \theta^0_b)-6 A(-\theta) \q .
\ea
We define the momenta to be
\ba\label{novemb2}
p_a^0=-\frac{\partial \tilde S_0}{\partial a_0} \, , \q\q\q  p_b=-\frac{\partial \tilde S_0}{\partial b_0}   \q .
\ea
On flat configurations we have 
\ba\label{novemb3}
{p_a^0}_{|flat}=3\frac{\partial A_a^0}{\partial a_0} (\pi -2\theta^0_a)+ 6\frac{\partial A^0_b}{\partial a_0}(\pi-2 \theta^0_b) \, , \q\q
{p_b^0}_{|flat}=6\frac{\partial A^0_b}{\partial b_0}(\pi-2 \theta^0_b) 
\ea
where for the  exterior angles we can write
\ba\label{novemb3a} 
\psi_a^0:=\pi-2\theta^0_a=-\pi+2\theta_a(a_0,b_0)\, , \q\q\q \psi_b^0=\pi-2\theta^0_b=-\pi+\theta_b(a_0,b_0) \q .
\ea
Here $\theta_a(a,b)$ and $\theta_b(a,b)$ are the dihedral angles in a simplex $\sigma(vijk\kappa)$ with $e(ij)=1,e(vi)=a$ and $e(v\kappa)=b$ at the triangles $\Delta(vij)$ and $\Delta(vi\kappa)$, respectively.   The reason for these relations is that the 3d $\text{star}(v_0)$ under consideration can be constructed by gluing two 4--simplices $\sigma(v_01234)$ and $\sigma(v_01235)$ together along their common tetrahedron $\tau(v_0 123)$. The 3d $\text{star}(v_0)$ is given by all the tetrahedra except for $\tau(1234),\tau(1235)$ and $\tau(v_0123)$. If we glue the triangulation corresponding to the tent move (with six 4--simplices) and these two 4--simplices along the 3d $\text{star}(v_0)$ together we can use flatness of the deficit angles at the triangles $\Delta(v_0ij)$ and $\Delta(v_0i\kappa)$ to conclude (\ref{novemb3}).

Hence, the momenta for flat configurations can be expressed as functions of the configuration variables $a_0,b_0$ only. Similarly, the components of the vector $Y^a,Y^b$ induced by displacing the vertex $v_0$ in the embedding flat spacetime can be expressed as functions of the configuration variables $a_0,b_0$ only. Note that due to our symmetry requirements there is just one such vector. A displacement of the vertex $v_0$ can only change the data associated to the star of $v_0$. For the complex of the two glued simplices it should {\it not} change the exterior angle $\psi_{\Delta(123)}$ at the triangle $\Delta(123)$ between the tetrahedra $\tau(1234)$ and $\tau(1235)$. This gives one relation between the two components of the vector $Y$ which can be computed to
\ba\label{novemb4}
Y^a= \frac{a_0^2-\tfrac{1}{3}}{a_0} \, , \q\q\q Y^b=\frac{a^2_0+b_0^2-1}{2b_0} \q .
\ea
and satisfies
\ba\label{novemb5}
Y^a\frac{\partial}{\partial a_0} \psi_{\Delta(123)}+ Y^b\frac{\partial}{\partial b_0} \psi_{\Delta(123)}=0 \q .
\ea
The linear constraint is according to (\ref{higher4}) the projection of the expression (\ref{novemb3}) for the flat momenta onto the vector $Y$, that is
\ba\label{novemb6}
C&=&Y^a\pi^0_a + Y^b\pi^0_b + \nn\\
&&\big(Y^a \frac{\partial }{\partial a_0} +Y^b \frac{\partial }{\partial b_0}\big)\bigg(  \big(\,\,3\frac{\partial A_a^0}{\partial a_0} (\pi-2\theta_a(a_0,b_0) ))+ 6\frac{\partial A^0_b}{\partial a_0}(\pi-\theta_b(a_0,b_0) )\,\,\big)\,\,y^a_0 +
\nn\\
&&\q\q\q\q\q\q\q\q \q \; 6\frac{\partial A^0_b}{\partial b_0}(\pi-\theta_b(a_0,b_0) ) )\,\, y^b_0   \q .
        \bigg)
\ea

\section{Constraint algebra}\label{algebra}

Gauge symmetries lead to constraints which in turn generate the gauge transformations in the canonical framework. The fact that the gauge transformations form a group is reflected in the first class property of the constraints, i.e.\ the Poisson bracket of two constraints should give a combination of constraints or vanish. 

In this section we will show that the linearized constraints (\ref{higher4}) are indeed Abelian and therefore first class. (Constraints that are linear and first class have to be Abelian.) As we will also consider constraints based at different vertices  $v,v'$ we will change the notation slightly, in particular we will omit the index for the time step $n$ and introduce an index $v$ for the constraints ${}^vC_I$ and the vectors ${}^vY^e_I$ based at the vertex $v$. We define ${}^vY_I^e=0$ if $e$ is not adjacent to $v$. 

We are considering an expansion around flat 4d space, hence we can assume that the triangulated hypersurface $\Sigma$ (with the background edge lengths $l^e$) is embedded into flat 4d space. Thus, there is some suitable flat 4d triangulation $\mathcal{T}$ such that $\Sigma$ is (part of) the boundary $\mathcal{B}(\mathcal{T})$ of $\mathcal{T}$. Furthermore, we will use the index $e$ for any edge in $\Sigma$, i.e.\ not only for the edges adjacent to $v$ or $v'$, and therefore not use the index $b$ anymore. By $\text{star}(v)$ we denote the 3d star of the vertex $v$ in $\Sigma$. Writing all summations over the index $e$ explicitly, the constraints (\ref{higher4}) are now
\ba\label{pb1}
{}^vC_I=\sum_{e\supset v} {}^vY_I^e \pi_e + \sum_{e'\subset \text{star}(v)} \sum_{e\supset v} {}^vY_I^e  \frac{\partial}{\partial l^e} \sum_{\Delta \subset \text{star}(v)} \frac{\partial A_{\Delta}}{\partial l^{e'}} (\pi-\theta_{\Delta}) y^e \q .
\ea
The Poisson bracket between two constraints is given by
\ba\label{pb2}
\{{}^vC_I, {}^{v'}C_J\}&=&\q
\sum_{{v'}\subset e \subset \text{star}(v)}\,\,\sum_{e' \supset v} \,\,{}^{v'}\!Y^e_J \,\, {}^v\!Y^{e'}_I 
\frac{\partial}{\partial l^{e'}} \sum_{\Delta \subset \text{star}(v)}\frac{\partial A_\Delta}{\partial l^e} (\pi- \theta_\Delta)
\nn\\
&&-
\sum_{{v}\subset e \subset \text{star}(v')}\,\,\sum_{e' \supset v'} \,\,{}^{v}\!Y^e_I \,\, {}^{v'}\!Y^{e'}_J
\frac{\partial}{\partial l^{e'}} \sum_{\Delta \subset \text{star}(v')}\frac{\partial A_\Delta}{\partial l^e} (\pi- \theta_\Delta) \q .
\ea

We will show that the two terms on the right hand side of (\ref{pb2}) cancel each other and hence the constraints commute. To  this end we will prove that both terms are second derivatives of
\ba\label{pb3}
S_{\mathcal{T}}:=\sum_{\Delta\subset \mathcal{B}(\mathcal{T})} A_{\Delta} (\pi- \theta_\Delta) +\sum_{\Delta\subset \mathcal{I}(\mathcal{T})}  A_\Delta \epsilon_\Delta
\ea
contracted with ${}^v\!Y_I^e$ and ${}^{v'}\!Y_J^e$. Here  $\Delta\subset \mathcal{B}(\mathcal{T}), \mathcal{I}(\mathcal{T}) $ denote triangles in the boundary and bulk of the 4d triangulation $\mathcal{T}$, respectively.  Using the Schl\"afli identity the first derivative evaluated on a flat configuration amounts to
\ba\label{pb4a}
\frac{\partial}{\partial l^e} S_{\mathcal{T}}{}_{|flat}=\sum_{\Delta\subset \mathcal{B}(\mathcal{T})}   \frac{\partial A_{\Delta}}{\partial l^e} (\pi- \theta_\Delta)  \q .
\ea
As the second derivative is contracted with a vector  ${}^v\!Y_I^e$ or  ${}^{v'}\!Y_J^e$ along which the configuration stays flat, we can still use the expression for the first derivative. Finally, note that 
\ba\label{pb4}
\sum_{e }\,\,\sum_{e' } \,\,{}^{v'}\!Y^{e'}_J \,\, {}^v\!Y^{e}_I 
\frac{\partial}{\partial l^{e'}}\,
% \sum_{\Delta \subset \text{star}(v)} 
\left( \frac{\partial A_\Delta}{\partial l^e} (\pi- \theta_\Delta)\right)
\ea
is only non--vanishing, if the triangle $\Delta$ is both in $\text{star}(v)$ and $\text{star}(v')$. Firstly, the derivative of the area $A_\Delta$ is only non--zero for $e\subset \Delta$, for the second derivative we additionally need $e'\subset \Delta$. Secondly, the derivative of the dihedral angle $\theta_\Delta$ is contracted with a vector that arises from displacing the vertex $v'$ in the flat 4d embedding space. Under such a displacement only dihedral angles associated to triangles in the star of $v'$ are affected (as only edges adjacent to $v'$ change, consequently, only the normals to the tetrahedra in the star of $v'$, and the normals determine the dihedral angles).  This shows that the second derivatives of $S_\mathcal{T}$ contracted with ${}^v\!Y_I^e$ and ${}^{v'}\!Y_J^e$ in the two possible ways give, indeed, the two terms in the Poisson bracket (\ref{pb2}), which therefore vanishes.

\section{Observables} \label{obs}

We have seen that in the canonical framework for linearized Regge calculus constraints appear. These have an Abelian algebra and, moreover, generate gauge transformations that correspond to translating the vertices of the hypersurface in the embedding flat (background) space time. Thus, not all variables are physical observables, that is invariant under gauge transformations. Gauge invariant variables are those that do not change under vertex translations and are associated to (linearized) curvature excitations. 

Considering the dynamical variables for a tent move at some $N$--valent vertex $v$ (we again omit the time step index $n$) we define a canonical transformation
\ba\label{ob1}
y^\Gamma &=&(T^{-1})^\Gamma_e\,\, y^e \nn\\
\pi_\Gamma &=& T^e_\Gamma \,\,\pi_e \q,
\ea
where we divide the index set $\Gamma=1,\ldots,N$ into two sets $I=1,\ldots,4$ and $\alpha=5,\ldots,N$, corresponding to gauge variables and gauge invariant variables, respectively.  For the $y^\alpha$ and $\pi_\alpha$ to commute with the constraints we need that
\ba\label{ob2}
Y^e_I(T^{-1})^\alpha_e =0 \,,\q\q  \q      Y^e_I N_{ee'}T^{e'}_\alpha  =0   \q .
\ea
Choosing $T^e_I=Y^e_I$ ensures the first equation in (\ref{ob2}), moreover, it then follows that the $\pi_I=Y_I^e\pi_e$ coincide with the momentum parts of the constraints. In particular, with the second condition in (\ref{ob2})
\ba\label{ob2a}
C_I&=& \pi_I + Y^e_I N_{ee'} Y^{e'}_{I'} \,y^{I'}+ Y^e_I N^n_{eb}\, y^b \q ,
\ea
that is the constraints only involve the gauge variables $y^I,\pi_I$.
%The matrix elements $(T^{-1})^\alpha_e$ can be found by choosing $N-4$ functions $f^\alpha$ that are invariant under translations of the vertex $v$ {\it( see below for examples)} and defining 
%\ba\label{ob3}
%(T^{-1})^\alpha_e=\frac{\partial}{\partial l^e} f^\alpha  \q .
%\ea
%Then the $y^\alpha$ correspond  to the linearizations of these invariant functions.

Partial and complete observables \cite{obs} are a general tool to compute physical observables and can also be applied here. Select four linearly independent edges (with respect to the geometry of the background solution) among the $N$ edges adjacent to $v$. We will split accordingly the index set $e=1,\ldots,N$ into $E=1,\ldots,4$ and $P=5,\ldots N$. The Poisson brackets between $y^E$ and the constraints give then an invertible matrix
\ba\label{ob3}
\{y^E,C_I\}=Y^E_I  \q .
\ea
(Note that the vectors $Y^e_I$ could have been chosen in a way such that $Y^E_I=\delta^E_I$, see the explanation in section \ref{deg}.) To any linear phase space function $f$ we can then associate an observable $F_f$ defined by
\ba\label{obs4}
F_f=f-\{f,C_I\}(Y^{-1})^I_E y^E  \q .
\ea
We can similarly proceed with any set of four momenta $\pi^E$, such that the matrix $\{\pi^E,C_I\}$ is invertible.
A geometrical interpretation of these observables will appear in \cite{hoehntoappear}.

In the following we will discuss the structure of the phase space for the dynamical variables of a tent move at an $N$--valent vertex $v$. As we have four constraints the constraint hypersurface is $(2N-4)$--dimensional. An $N$--dimensional submanifold of this hypersurface is given by configurations (here linearized length and momentum variables) leading to flat geometries. For these configurations all the momenta are fixed as functions of the length variables -- the relations can be obtained by linearizing the formula for the momenta $\tilde p^e$ valid for flat geometries (\ref{higher1}). Furthermore, we have 4--dimensional gauge orbits in the constraint hypersurface. Note that these gauge transformations also leave the subspace of flat configurations invariant.  Given a point $p$ in the subspace of flat configurations there are $(N-4)$ directions transversal to the gauge orbits but tangential to this subspace, i.e.\ leading to flat configurations which are not in the gauge orbit of the point $p$. There are another $(N-4)$ directions transversal to the subspace of flat configurations but inside the constraint hypersurface. These directions lead to geometries with (linearized) curvature.

These constructions can be enlarged to apply to all constraints  at all the vertices of $\Sigma$ \cite{hoehntoappear}. For the counting of gauge invariant variables note that not all constraints are linearly independent. If one considers, for example, the boundary of a 4--simplex as a 3d hypersurface $\Sigma$ one counts four constraints at each of the five vertices, hence 20 constraints. Only ten of these are linearly independent, however, since the other ten generate global translations and 4d rotations of the simplex in the embedding 4d flat space. As there are only ten edge variables in $\Sigma$ the physical phase space is zero--dimensional.

\subsection{Example: five--valent symmetry--reduced vertex}

Here we consider again the five--valent symmetry--reduced vertex from section \ref{beispiel1}. Using an auxilary construction of the 3d $\text{star}(v)$ via the gluing of two 4--simplices $\sigma(v1234)$ and $\sigma(v1235)$, we already observed in section (\ref{beispiel1}) that the exterior angle $\psi_{\Delta(123)}$ at the triangle $\Delta(123)$ is invariant under displacements of the vertex $v$, that is
\ba\label{cc1}
(Y^a \frac{\partial}{\partial a} + Y^b \frac{\partial }{\partial b} ) \psi_{\Delta(123)}=0 \q .
\ea
Thus, 
\ba\label{cc2}
y^{\psi}:= \frac{\partial   \psi_{\Delta(123)}  }{\partial a}y^a + \frac{\partial   \psi_{\Delta(123)}  }{\partial b} y^b
\ea
is an observable of the linearized theory. It is the unique observable linear in the configuration variables $y^a,y^b$ (modulo rescaling). The dihedral angle $\psi_{\Delta(123)}$ would be the only part of the deficit angles in the bulk which depends on the lengths adjacent to the vertex $v$. Hence, the linearized observables are related to the (linearized) deficit angles and ultimately to the identities (\ref{bia2}), expressing the invariance of the deficit angles under the vector fields $Y_I$ evaluated on a flat background. This invariance, for instance for the deficit angles hinging at the tent pole, can be confirmed numerically in this example.

Explicitly, $\psi_{\Delta(123)}$ is given by
\ba
\psi_{\Delta(123)}={\rm{arcsec}}\left(\frac{2\sqrt{6a^2-2}}{3a^2-3b^2+1}\right) \, ,
\ea
so that 
\ba\label{5v2}
y^{\psi}
&=&\frac{1}{\sqrt{-3a^4-3b^4+6a^2b^2+6a^2+2b^2-3}} \left(  -\frac{3\sqrt{3}a(a^2+b^2-1)}{(3a^2-1)}   \, y^a 
 +2\sqrt{3}b\, y^b \right)  \,  .\q\q
\ea
The Poisson bracket of $y^\psi$ with the constraint $C$ in (\ref{novemb6}) can be explicitly computed and is vanishing.

Next, we will construct the matrix $T$ and in this way also obtain a momentum observable. One choice for $T$ is to define (assuming the generic case $\text{det}(N)\neq0$)
\ba\label{cc3}
N_{gauge\,\, a}:=Y^a N_{aa}+ Y^b N_{ba} \, \q\q\q  N_{gauge\,\, b}:=Y^a N_{ab}+ Y^b N_{bb} 
\ea 
and 
\ba\label{cc4}
{T_\Gamma}^e=
\left(
\begin{array}{cc}
Y^a & Y^b \\
-N_{gauge\,\,b} & N_{gauge\,\, a}
\end{array}\right) \q ,
\ea
where the indices take values $\Gamma=\{gauge,\, obs\}$ and $e=\{a,b\}$. It clearly satisfies the conditions in (\ref{ob2}). The inverse is then proportional to
\ba\label{cc5}
{(T^{-1})_e}^\Gamma=
\left(
\begin{array}{cc}
N_{gauge\,\,a} & -Y^b \\
N_{gauge\,\,b} & Y^a
\end{array}\right)  \q .
\ea

The observable $y^{obs}=\sum_e y^e{(T^{-1})_e}^{obs}$ is proportional to $y^\psi$ as defined by (\ref{cc2}). The constraint (ignoring boundary variables) can now be expressed as
\ba\label{cc6}
C=\pi_{gauge}+ N_{gauge\,\,gauge} \, y^{gauge}
\ea
where $N_{gauge\,\,gauge}=\sum_{e,e'}Y^eN_{ee'}Y^{e'}$. 

The explicit expressions are quite lengthy, but can be computed in a straightforward way. For instance, for the specific configuration $a=1,b=1$ we obtain
\ba\label{cc7}
y^{\psi}=\sqrt{\frac{3}{5}}\left(-\frac{3}{2}\, y^a +2\,  y^b\right) \q,
\ea
while the momentum observable, defined via (\ref{ob1}), reads
\ba\label{cc8}
\pi_{obs}=-0.548142\, \pi_a -0.85287\, \pi_b \q.
\ea
Finally, the constraint (\ref{cc6}) is 
\ba\label{cc9}
C=\pi_{gauge}-0.294509\, y^{gauge}=0\q.
\ea

\section{The dynamics of gravitons as generated by the tent moves}\label{dyn}

The constraints (\ref{lin6}) are linear in the perturbation variables. Besides restricting the allowed dynamical configuration, the constraints also generate the changes in the edge lengths and momenta induced by infinitesimal displacements of the vertices. However, they leave observables - corresponding to the graviton degrees of freedom - invariant, i.e.\ do not generate any dynamics for these. In the continuum the dynamics of the linearized theory with respect to the background time (i.e.\ the time defined by the flat solution and a $3+1$ decomposition of this solution) is generated by a quadratic, global Hamiltonian. Here the dynamics is described by tent moves. Also for the discretized case this dynamics is defined with respect to the time as defined by the background solution: We evolve the configuration in discrete steps and the proper distance between the vertices in different time steps is essentially determined by the background solution.

The tent move dynamics for the perturbation variables is given by the equations (\ref{lin4})
\ba\label{dyn1}
\pi_e^n&=& 
- M_{ee'}^n y_{n+1}^{e'} 
 -  N_{ee'}^n y_n^{e'} 
 - N_{eb}^n y^b\q ,
\ea
where the matrices appearing in (\ref{dyn1}) have been defined in (\ref{lin5}).

The momenta at the next time step $\pi_e^{n+1},\pi_b^{n+1}$ are defined by (\ref{oct3}, \ref{zus3}), the linearization of which  gives 
\ba\label{sep16}
\pi_e^{n+1} &=&\q\q
 y_n^{e'}M_{e'e}^n
+ {N'}_{ee'}^n y_{n+1}^{e'}
+ {N'}_{eb}^n y^b \nn\\
\pi_b^{n+1}&=&\pi^n_b\,+\, N^n_{be}y^e_n + {N'}^n_{be}y^e_{n+1}+ N_{bb'}^n y^{b'} \q .
\ea
Here we additionally introduced the matrices 
\ba\label{sep18}
{N'}_{ee'}^n &=&    \frac{\partial^2 S_n}{\partial l^e_{n+1} \partial l^{e'}_{n+1}}  -     \frac{\partial^2 S_n}{\partial l^e_{n+1} \partial t_{n}} 
\left(  \frac{\partial^2 S_n}{\partial t_{n} \partial t_{n}} 
  \right)^{-1}   
   \frac{\partial^2 S_n}{\partial t_{n} \partial l^{e'}_{n+1}}
          \nn\\
{N'}_{eb}^n&=& 
  \frac{\partial^2 S_n}{\partial l^e_{n+1} \partial l^{b}}  -     \frac{\partial^2 S_n}{\partial l^e_{n+1} \partial t_{n}} 
\left(  \frac{\partial^2 S_n}{\partial t_{n} \partial t_{n}} 
  \right)^{-1}   
   \frac{\partial^2 S_n}{\partial t_{n} \partial l^{b}} 
\ea
and
\ba\label{sep18b}
    {N}_{be}^n&=& 
  \frac{\partial^2 S_n}{\partial l^b\partial l^{e}_n}  -     \frac{\partial^2 S_n}{\partial l^b \partial t_{n}} 
\left(  \frac{\partial^2 S_n}{\partial t_{n} \partial t_{n}} 
  \right)^{-1}   
   \frac{\partial^2 S_n}{\partial t_{n} \partial l^{e}_n} \nn\\ 
     {N'}_{be}^n&=& 
  \frac{\partial^2 S_n}{\partial l^b \partial l^{e}_{n+1}}  -     \frac{\partial^2 S_n}{\partial l^b \partial t_{n}} 
\left(  \frac{\partial^2 S_n}{\partial t_{n} \partial t_{n}} 
  \right)^{-1}   
   \frac{\partial^2 S_n}{\partial t_{n} \partial l^{e}_{n+1}}  \nn\\
    {N}_{bb'}^n&=& 
  \frac{\partial^2 S_n}{\partial l^b\partial l^{b'}}  -     \frac{\partial^2 S_n}{\partial l^b\partial t_{n}} 
\left(  \frac{\partial^2 S_n}{\partial t_{n} \partial t_{n}} 
  \right)^{-1}   
   \frac{\partial^2 S_n}{\partial t_{n} \partial l^{b'} }   \q .
\ea
The equations (\ref{dyn1}) have to be used to determine the variables $y_{n+1}^e$ as a function of the momenta $\pi_e^n$ and the variables $y_n^e,y^b_n$. The variables $y^e_{n+1}$ at time step $(n+1)$ can, however, not be uniquely determined as $M_{ee'}^n$ is not invertible and as shown in section \ref{deg} has four right null vectors ${}^{n+1}\!{Y}^e_I$. (Here we introduced an additional index $(n+1)$ as these are the null vectors defined by the background geometry at time step $(n+1)$. The left null vectors  of $M^n_{ee'}$ are ${}^nY^e_I$ which we so far denoted by just $Y^e_I$.)  This non--uniqueness reflects the gauge freedom in the evolution. We will use the splitting of the variables into gauge variant and gauge invariant variables introduced  in (\ref{ob1}) 
\ba\label{sep19}
y_{n}^e  &=& {}^{n}\!T_\alpha^e y^\alpha_{n} \,\,+\,\, {}^{n}\!T_I^e y^I_{n} \nn\\
\pi^{n}_e  &=& {}^{n}\!(T^{-1})^\alpha_e \pi_\alpha^{n} \,\,+\,\, {}^{n}\!(T^{-1})^I_e \pi_I^{n} \nn\\
y_{n+1}^e  &=& {}^{n+1}\!T_\alpha^e y^\alpha_{n+1} \,\,+\,\, {}^{n+1}\!T_I^e y^I_{n+1} \nn\\
\pi^{n+1}_e  &=& {}^{n+1}\!(T^{-1})^\alpha_e \pi_\alpha^{n+1} \,\,+ \,\,{}^{n+1}\!(T^{-1})^I_e \pi_I^{n+1} 
 \q .
\ea
Equation (\ref{dyn1}) relating the momenta $\pi^n_e$ and the length variables $y^{n+1}_e$ becomes
\ba\label{dyn2}
{}^{n}\!(T^{-1})^\alpha_e \pi_\alpha^{n} \,\,+\,\, {}^{n}\!(T^{-1})^I_e \pi_I^{n}&=&
-M^n_{ee'} \left(      {}^{n+1}\!T_\alpha^{e'} y^\alpha_{n+1} \,\,+\,\, {}^{n+1}\!T_I^{e'} y^I_{n+1}           \right) \nn\\
&&-N^n_{ee'}\left(    {}^{n}\!T_\alpha^{e'} y^\alpha_{n} \,\,+\,\, {}^{n}\!T_I^{e'} y^I_{n}    \right) \,-\,N^n_{eb}y^b\q .
\ea
Multiplying this equation with ${}^n\!T^e_\alpha$ and remembering that ${}^n\!T^e_I= {}^n\!Y^e_I$ and ${}^{n+1}\!T^e_I= {}^{n+1}\!Y^e_I$ we obtain
\ba\label{dyn3}
\pi^n_{\alpha'}&=& -  \left( {}^{n}\!T^{e}_{\alpha'} M^n_{ee'}\, {}^{n+1}\!T_\alpha^{e'}\right) y^\alpha_{n+1}
 -\left( {}^{n}\!T^{e}_{\alpha'} N^n_{ee'}\, {}^{n}\!T_\alpha^{e'}\right) y^\alpha_{n}
 -\left( {}^{n}\!T^{e}_{\alpha'} N^n_{eb} \right) y^b\q ,
\ea
where we used the conditions (\ref{ob2}) on the transformation matrix $T$. Now the null vectors of $M^n_{ee'}$ are projected out and we can invert equation (\ref{dyn3}) for the invariant combinations of the length variables $y^\alpha_{n+1}$. The $y^I_{n+1}$ are left undetermined and, consequently, can be freely chosen. These variables correspond therefore to lapse and shift. Similarly to equation (\ref{dyn3}), one can show
\ba\label{dyn4}
\pi^{n+1}_{\alpha'}&=&   y^\alpha_{n}
  \left( {}^{n}\!T^{e}_{\alpha'} M^n_{ee'}\, {}^{n+1}\!T_\alpha^{e'}\right) 
   +\left( {}^{n+1}\!T^{e}_{\alpha'} {N'}^n_{ee'}\, {}^{n+1}\!T_\alpha^{e'}\right) y^\alpha_{n+1}
 +\left( {}^{n+1}\!T^{e}_{\alpha'} {N'}^n_{eb} \right) y^b \,.\q
\ea
To this end one has to confirm that 
\ba\label{dyn5}
 {}^{n+1}\!T^{e}_{\alpha'} {N'}^n_{ee'}\, {}^{n+1}Y^{e'}_I=0
\ea
which follows from the condition (\ref{ob2}) on the transformation matrix ${}^{n+1}\!T$
\ba\label{dyn6}
 {}^{n+1}\!T^{e}_{\alpha'} {N}^{n+1}_{ee'}\,\, {}^{n+1}Y^{e'}_I=0
\ea
and the fact that ${N}^{n+1}_{ee'}\, {}^{n+1}Y^{e'}_I=-{N'}^n_{ee'}\, {}^{n+1}Y^{e'}_I$, derived in equation (\ref{oct10a}). 

With (\ref{dyn3}) and (\ref{dyn4}) we reduced the dynamics onto the gauge invariant variables.

In the rest of this section we will show  that the constraints at the next time step are automatically satisfied. To show the preservation of constraints at the vertex $v$ itself we contract the equation (\ref{sep16}) 
for the momenta $\pi^{n+1}_e$ 
\ba\label{dyn7}
\pi_e^{n+1} &=&
 y_n^{e'}M_{e'e}^n
+ {N'}_{ee'}^n y_{n+1}^{e'}
+ {N'}_{eb}^n y^b
\ea
with the vector fields ${}^{n+1}Y^e_I$. Using again ${N}^{n+1}_{ee'}\, {}^{n+1}Y^{e'}_I=-{N'}^n_{ee'}\, {}^{n+1}Y^{e'}_I$ from equation (\ref{oct10a}) and ${N}^{n+1}_{ee'}\, {}^{n+1}Y^{e'}_I=-{N'}^n_{ee'}\, {}^{n+1}Y^{e'}_I$ from equation (\ref{oct10c}) we obtain
\ba\label{dyn8}
{}^{n+1}\!Y^e_I\pi^{n+1}_e&=& - {}^{n+1}\!Y^e_I N^{n+1}_{ee'}y^{e'}_{n+1} -  {}^{n+1}\!Y^e_I N^{n+1}_{eb}y^{b}\q ,
\ea
that is the constraints (\ref{lin6}) at time step $(n+1)$.
%% S(\alpha,I)=S(\alpha)%%%then I variables drop out completely

\begin{figure}[hbt!]
\begin{center}
    \psfrag{v}{$v_{n-1}=v_n$}
    \psfrag{vp}{$v'_{n-1}$}
    \psfrag{vpn}{$v'_n=v'_{n+1}$}
    \psfrag{vn}{$v_{n+1}$}
    \psfrag{S}{$\Sigma_{n-1}$}
    \includegraphics[scale=0.4]{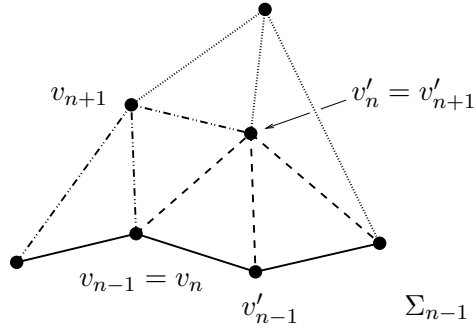}
    \end{center}
    \caption{\label{neitm}{\small Schematic illustration of the alternating tent moves at neighbouring vertices $v$ and $v'$, starting from the Cauchy hypersurface $\Sigma_{n-1}$.}}
\end{figure}

We also have to show that the constraints at the neighbouring vertices $v'$ are satisfied after a tent move at $v$ has been performed. This is slightly more involved, however, quite straightforward if one starts from the covariant picture. To this end, consider a specific neighbouring vertex $v'$ and the situation schematically represented in figure \ref{neitm}. 
That is, at a time $(n-1)$ we firstly perform a tent move at $v'_{n-1}$, then a tent move at $v_n=v_{n-1}$ and, finally, at $v'_{n}=v'_{n+1}$.  The action for these three tent moves is $S=S_{n-1}+S_n+S_{n+1}$ associated to a piece of triangulation $\mathcal T$ with boundary and one bulk vertex $v'_n=v'_{n+1}$.  With $\tilde S$ we denote the action with the tent pole edges integrated out. To simplify notation we use $E$ as an index for all edges in this triangulation (including the boundary) except for the edges $\overline{e}_n=e(v'_n v_{n})$ and $\overline{e}_{n+1}=e(v'_n v_{n+1})$.  With $E'$ we denote all edges adjacent to $v'_n=v'_{n+1}$ except for the edges $\overline{e}_n=e(v'_n v_{n})$ and $\overline{e}_{n+1}=e(v'_n v_{n+1})$.  The indices $A,A'$ run over all edges of the triangulation $\mathcal{T}$.

We have one inner vertex in the triangulation and hence four null vectors ${}^{v'}Y_I$ for the Hessian whose components are ${}^{v'}Y_I^{E'},{}^{v'}Y_I^{\overline{e}_n},{}^{v'}Y_I^{\overline{e}_{n+1}}$. Note that
\ba\label{fine1}
0&=&{}^{v'}Y_I^{E'} \frac{\partial^2 \tilde S}{l^{E'}l^A} + {}^{v'}Y_I^{\overline{e}_n} \frac{\partial^2 \tilde S}{l^{\overline{e}_n}   l^A} 
+ {}^{v'}Y_I^{\overline{e}_{n+1}} \frac{\partial^2 \tilde S}{l^{\overline{e}_{n+1}}   l^A}
\ea
holds for all edges $A$ in the triangulation, also for edges in the boundary. To see this either apply arguments similar to those in section \ref{deg} or, alternatively, realize that the ${}^{v'}Y_I$ would also be null vectors for the Hessian associated to a bigger triangulation containing the triangulation $\mathcal T$. But the first derivative in (\ref{fine1}) is with respect to edges adjacent to $v'_n$, hence even for the bigger triangulation there is only the action $\tilde S$ associated to $\mathcal T$ involved.

To obtain the constraints at $v'_n$, we firstly contract the equations (\ref{fine1}) with the variables $y^A$. Secondly, we use the definition for the momenta at $v'_n$
\ba\label{fine2}
\pi^{E'}_n= \frac{\partial^2 \tilde S_{n-1}}{l^{E'} l^A} y^A \q,  \q\q\q  \pi^{\overline{e}_n}= \frac{\partial^2 \tilde S_{n-1}}{l^{\overline{e}_n} l^A} y^A
\ea
and thirdly the equation
\ba\label{fine3}
{}^{v'}Y_I^{\overline{e}_{n+1}}=- {}^{v'}Y_I^{E'}       \frac{\partial^2 \tilde S  } {\partial l^{E'}  \partial  l^{\overline{e}_{n+1}}  }
\left( 
\frac{\partial^2 \tilde S  } {\partial l^{\overline{e}_{n+1}}    \partial  l^{\overline{e}_{n+1}}  } \right)^{-1}
-{}^{v'}Y_I^{\overline{e}_{n}}     \frac{\partial^2 \tilde S  } {\partial l^{\overline{e}_{n}} \partial  l^{\overline{e}_{n+1}}  }
\left( 
\frac{\partial^2 \tilde S  } {\partial l^{\overline{e}_{n+1}}    \partial  l^{\overline{e}_{n+1}}  }
\right)^{-1}    \q
\ea
between the components of ${}^{v'}Y_I$ that follows from the equations (\ref{fine1}) with $A$ set to $\overline{e}_{n+1}$. We obtain
\ba\label{fine4}
0&=&
{}^{v'}Y_I^{E'}  \pi_{E'}^n  + {}^{v'}Y_I^{\overline{e}_{n}} \pi_{\overline{e}_{n}} + \nn\\&&
{}^{v'}Y_I^{E'}
 N^{n,n+1}_{E'E}
  y^E+
{}^{v'}Y_I^{\overline{e}_{n}}  
N^{n,n+1}_{\overline{e}_n E}
   y^E +
     {}^{v'}Y_I^{E'}   N^{n,n+1}_{E' \overline{e}_n} y^{\overline{e}_n}+
  {}^{v'}Y_I^{\overline{e}_{n}}   N^{n,n+1}_{\overline{e}_n   \overline{e}_n} y^{\overline{e}_n} +
  \nn\\&&
   {}^{v'}Y_I^{E'}   N^{n,n+1}_{E' \overline{e}_{n+1}}  y^{\overline{e}_{n+1}} +
     {}^{v'}Y_I^{\overline{e}_{n}}    N^{n,n+1}_{{\overline{e}_{n}} \overline{e}_{n+1}}  y^{\overline{e}_{n+1}} \q ,
\ea
where $N^{n,n+1}_{A A'}$ is given by
\ba
N^{n,n+1}_{AA'}=
 \left(    \frac{\partial^2 (\tilde S_{n}+ \tilde S_{n+1})}{l^{A} l^{A'}}      -      \frac{\partial^2 \tilde S  } {\partial l^{A}  \partial  l^{\overline{e}_{n+1}}  }
\left( 
\frac{\partial^2 \tilde S  } {\partial l^{\overline{e}_{n+1}}    \partial  l^{\overline{e}_{n+1}}  } \right)^{-1}    
   \frac{\partial^2 \tilde S  } {  \partial  l^{\overline{e}_{n+1}}  \partial l^{A'} }   \right)
\ea
and coincides with the Hessian of the `effective action' obtained by integrating out the length associated to the edge $\overline{e}_{n+1} $ from $\tilde S_{n}+\tilde S_{n+1}$. This explains also the vanishing of $ N^{n,n+1}_{A \overline{e}_{n+1}}$ in the last line of (\ref{fine4}). Hence, these equations do not depend on the variable $y^{\overline{e}_{n+1}}$ and one can check that (\ref{fine4}) indeed reduce to the constraints  at $v'_n$ at time $n$.  To this end, note that according to (\ref{fine1})
\be\label{fine5}
{}^{v'}Y_I^{E'}   N^{n,n+1}_{E' A} +  {}^{v'}Y_I^{\overline{e}_{n}}    N^{n,n+1}_{{\overline{e}_{n}}  A} =-    {}^{v'}Y_I^{E'}  
\frac{\partial^2 \tilde S_{n-1}  } {\partial l^{E'}    \partial  l^A  } - {}^{v'}Y_I^{\overline{e}_{n}}   \frac{\partial^2 \tilde S_{n-1}  } {\partial l^{\overline{e}_{n}}    \partial  l^A  } 
\ee
which shows that the constraint equations (\ref{fine4}), indeed, involve only variables from time step $n$. 

Next, we want to obtain the constraints at the vertex $v'_{n+1}$ at time step $(n+1)$ starting from the constraints at time step $n$.  We use the defining equations for the dynamics of the tentmove (\ref{dyn1}, \ref{sep16})
\ba\label{fine6}
\pi^{E'}_{n+1} = \pi^{E'}_n+  \frac{\partial^2 \tilde S_{n}}{l^{E'} l^A} y^A \q ,  \q\q  \pi^{\overline{e}_{n+1}}= \frac{\partial^2 \tilde S_{n}}{l^{\overline{e}_{n+1}} l^A} y^A\q,\q\q   \pi^{\overline{e}_{n}}= -\frac{\partial^2 \tilde S_{n}}{l^{\overline{e}_{n}} l^A} y^A
\ea
and, furthermore, the relation (\ref{fine3}) between the components of the null vectors ${}^{v'}Y_I$ in `backward direction'. In the resulting expression all terms involving ${}^{v'}Y^{\overline{e}_n}_I$ or $y^{\overline{e}_n}$ cancel each other and we  obtain
\ba\label{fine7}
0&=&
{}^{v'}Y_I^{E'}  \pi_{E'}^{n+1}  + {}^{v'}Y_I^{\overline{e}_{n+1}} \pi_{\overline{e}_{n+1}} + \nn\\&&
{}^{v'}Y_I^{E'} N^{n+1}_{E'E} y^E+
{}^{v'}Y_I^{\overline{e}_{n+1}}  
N^{n+1}_{\overline{e}_{n+1} E}  y^E +
     {}^{v'}Y_I^{E'}   N^{n+1}_{E' \overline{e}_{n+1}} y^{\overline{e}_{n+1}}+
  {}^{v'}Y_I^{\overline{e}_{n+1}}   N^{n}_{\overline{e}_{n+1}   \overline{e}_{n+1}} y^{\overline{e}_{n+1}}  \q , \q\q\q
\ea
where
\be\label{fine8}
N^n_{AA'}=\frac{\partial^2 \tilde S_n}{\partial l^A \partial l^{A'}} \q .
\ee
The equations (\ref{fine7}), in fact, constitute the constraints at $v'_{n+1}$. Therefore, the tent move dynamics at the vertex $v$ preserves the constraints also at the neighbouring vertices $v'$. 

The (commutation) algebra of tent moves at neighbouring vertices will be considered in further work \cite{hoehntoappear,ta2}, as well as the generalization of a dynamics defined by tent moves to a dynamics defined by Pachner moves \cite{dh2}.

\subsection{Example: symmetry--reduced five--valent vertex}

Using the splitting into gauge invariant and gauge variant variables, the dynamics completely decouples. At each time step we have a constraint
\ba\label{cc10}
C=\pi_{gauge}^n+ N_{gauge\,\,gauge}^n \, y^{gauge}_n
\ea
fixing the gauge momentum as a multiple of the gauge variable $y^{gauge}_n$, which, on the other hand, can be freely chosen. The gauge invariant variables at different times are coupled through
\ba\label{cc11}
\pi^n_{obs} &=& -M_{obs\,\, obs}^n\, y^{obs}_{n+1} - N^n_{obs\,\, obs} \,y^{obs}_n \nn\\
\pi^{n+1}_{obs}&=&y^{obs}_n \,M_{obs\,\, obs}^n +{N'}^n_{obs\,\, obs}\, y^{obs}_{n+1} 
\ea
where $M_{obs\,\,obs}^n =\sum_{ee'} T^e_{obs} M_{ee'}^n  T^{e'}_{obs}$ and so on.

Specifically evaluating the last equation for the first two time steps on the flat background with $a_0=b_0=1$ and $t_0=1/10$ yields
\ba\label{cc12}
\pi^0_{obs} &=& -44.1584 \, y^{obs}_{1} -36.3356 \,y^{obs}_0 \nn\\
\pi^{1}_{obs}&=&44.1584 \, y^{obs}_0 +31.0145\, y^{obs}_{1} \q.
\ea

\section{Higher order dynamics and pseudo constraints}\label{higher}

In the previous sections we discussed the theory defined by an expansion of the action to quadratic order on the flat background. Note that this background solution also displays gauge symmetries, namely translations of the vertices. For the linearized theory we obtained constraints -- arising as equations of motion which only depend on the dynamical data associated to one time step. Although the constraints could have been dependent on the background gauge parameters at the next or previous time steps (which do not belong to the dynamical variables), they actually did not. In fact, if the constraints were dependent on the background gauge parameters at the other time steps, it would have been impossible to obtain consistency of the constraint evolution (assuming local evolution laws).

However, we expect gauge symmetries to be broken for the full non--linear theory \cite{bahrdittrich09a}. Hence, starting with some higher order of the expansion of the Regge action, the gauge freedom should become fixed by the equations of motion. Here the lowest order to become fixed is actually not the first order variables but the background gauge itself. Interestingly, this fixing of the background gauge minimizes the dependence of the Hamilton principal function (sometimes called the Hamilton--Jacobi function), i.e.\ the action evaluated on the solution as a function of the boundary data, on the background gauge. 

In the remainder of this section we will discuss this mechanism, which will turn the constraints into pseudo constraints, i.e.\ equations of motion which depend on lapse and shift \cite{friedmann, gambini}.

Firstly, we will discuss the covariant formulation. The canonical description can be obtained afterwards as a rewriting of the equations of motion. In order to simplify the formulae, we will ignore variations $y^b$ of the edges in the boundary of the tent moves. We consider two consecutive tent moves from time step $n=0$ to time step $n=2$ and consider a boundary value problem with data given for times $n=0,2$ and free variables at time $n=1$. Moreover, we will assume that the lengths of the tent pole edges $t_0,t_1$ have been integrated out, that is we will work with the effective actions $\tilde S_0,\tilde S_1$.

We expand the length variables as 
$l^e_n={}^fl^e_n+\varepsilon\, \,{}^{(1)}y^e_n + \varepsilon^2 \,\, {}^{(2)}y^e_n+ O(\varepsilon^3)$ and proceed similarly for the momenta. Furthermore, we will use the split of the variables into gauge variant and gauge invariant ones defined by the linearized theory, i.e.\ we will use the transformation matrix ${}^{n}\!T^e_\Gamma$ satisfying the conditions in (\ref{ob2}).

The equations of motion (contracted with ${}^1\!T^e_\Gamma$),
\ba\label{dd1}
0&=&\frac{\partial (\tilde S_0+ \tilde S_1)}{\partial l^e_1} \ {}^1\!T^e_\Gamma\q ,
\ea
expanded to second order are given by
\ba\label{dd2}
0 &=&     \sum_{n=0,1,2} \frac{\partial^2 (\tilde S_0+ \tilde S_1)}{\partial l^e_1 \partial l^{e'}_n}  \,\,{}^1\!T^e_\Gamma\,\, {}^n\!T^{e'}_{\Gamma'} \,\, \left(\varepsilon \, {}^{(1)}\!y^{\Gamma'}_n + \varepsilon^2 \,{}^{(2)}\!y^{\Gamma'}_n\right)   \nn\\
  &&+ \tfrac{1}{2}\sum_{n,n'=0,1,2}             
      \frac{\partial^3 (\tilde S_0+ \tilde S_1)}{\partial l^e_1 \partial l^{e'}_n   \partial l^{e''}_{n'}} 
       \,\,{}^1\!T^e_\Gamma\,\, {}^n\!T^{e'}_{\Gamma'}     \,\,   {}^{n'}\!T^{e'}_{\Gamma''}   \,\,   \varepsilon^2 
        \, {}^{(1)}\!y^{\Gamma'}_n\, \, {}^{(1)}\!y^{\Gamma''}_{n'}   \q .
\ea

In the following we will consider the equations from the variation of the gauge variables, i.e.\ equations with index $\Gamma=I$. For these the first line in (\ref{dd2}) vanishes as it contains the Hessian of the action contracted with the null vector ${}^{1}\!Y^e_I={}^{1}\!T^e_I$. Denoting the second order terms by $\mathfrak S_I$ we will show the following \\
~\\
{\bf Claim:}  $\mathfrak S_I$ coincides with the derivative of Hamilton's principal function truncated to second order in the direction of the null vectors ${}^1Y_I$.\\
~\\
{\bf Proof:}  The proof will proceed in two main steps. First we will show that -- if we use the linearized equations of motion -- all terms with gauge variables in $\mathfrak S_I$ vanish. Consequently, there are no variables left to solve for and we have to use equation (\ref{dd2}) as a consistency equation for the background gauge at $n=1$. In a second step, we will show that $\mathfrak S_I$ coincides with the derivative of the second order Hamilton's principal function with respect to the background gauge.

To begin with, we introduce the notation
\ba\label{dd7}
(I_1 \Delta'_{n'} \Delta''_{n''}):=\tfrac{1}{2}\sum_{\Delta',\Delta''}   \frac{\partial^3 (\tilde S_0+ \tilde S_1)}{\partial l^e_{1} \partial l^{e'}_{n'}  \partial l^{e''}_{n''}} 
       \,\,{}^1\!T^e_I\,\, {}^{n'}\!T^{e'}_{\Delta'}     \,\,   {}^{n''}\!T^{e''}_{\Delta''}   \,\,   \varepsilon^2 
        \, {}^{(1)}\!y^{\Delta'}_{n'}\, \, {}^{(1)}\!y^{\Delta''}_{n''}  \q 
\ea
where $\Delta',\Delta''$ can stand for the gauge indices $I',I''$ or for the obsevable indices $\alpha',\alpha''$.
~\\
We start by showing that\\
~\\
{\bf (a)} all terms $(I_1 I'_{n'}  \Delta''_{n''} )$ with $n'\neq1$ and all terms $(I_1 \Delta'_{n'}  I''_{n''} )$ with $n''\neq 1$ vanish.\\
~\\
Consider, for instance, $(I_1 I'_0 \Delta''_{n''})$. The third derivative appearing in this term can be rewritten as
\ba\label{dd3}
  \frac{\partial^3 \tilde S_0}{\partial l^e_1 \partial l^{e'}_0   \partial l^{e''}_{n'}} 
       \,\,{}^1\!T^e_I\,\, {}^0\!T^{e'}_{I'}  = {}^{1}\!Y^e_I   \frac{\partial}{\partial l^e_1} \left( {}^{0}\!Y^{e'}_{I'}    \frac{\partial}{\partial l_0^{e'}  }           \frac{\partial \tilde S_0}{\partial l^{e''}_{n'} }    \right) \q ,
\ea 
where we used that ${}^{0}\!Y^{e'}_{\Gamma'}$ can be expressed as functions of the length variables $l^{e}_0$ only (and lengths in the boundary of the tent). For $n'=2$ the expression in (\ref{dd3}) vanishes. For $n'=0,1$ we can understand the right hand expression as the double derivative of the momentum at time $n'=0$ or at time $n'=1$, respectively. Since both derivatives are in flat directions in configuration space, we can use  the expression for 
\be\label{dd4}
\tilde p_{e''}^n {}_{|flat}   =  (-1)^{n-1}   \frac{\partial \tilde S_0} {\partial l^{e''}_{n'}}  \,\,{}_{|flat}
\ee
which is valid on flat configurations. On this subspace of the configuration space $\tilde p^n_{e''}$ is a function of the variables $l^e_n$ only, hence either the derivative with respect to $l^{e'}_0$ or the derivative with respect to $l^e_1$ will force the expression in (\ref{dd3}) to vanish. 

In conclusion, all second order terms in the equations of motion associated to the gauge index $I$ which contain gauge variables at times $n=0$ or $n=2$ vanish. Note that these terms would also vanish if we considered the second order momenta.

~\\
Next, we show that\\
~\\
{\bf (b)} all terms $(I_1 I'_{1}  \Delta''_{n''} )$ and all terms $(I_1 \Delta'_{n'}  I''_{1} )$ vanish if one uses the first order equations of motions for the ${}^{(1)}\!y^\alpha_1$ .\\
~\\
We use a similar rewriting as in {\bf (a)}, that is 
\ba\label{dd5}
 \frac{\partial^3 (\tilde S_0+ \tilde S_1)}{\partial l^e_1 \partial l^{e'}_1   \partial l^{e''}_{n'}} 
       \,\,{}^1\!T^e_I\,\, {}^1\!T^{e'}_{I'}    = 
        {}^{1}\!Y^e_I   \frac{\partial}{\partial l^e_1} \left( {}^{1}\!Y^{e'}_{I'}    \frac{\partial}{\partial l_1^{e'}  }   
                \frac{\partial (\tilde S_0 + \tilde S_1)}{\partial l^{e''}_{n'} }    \right) - 
               \left(    {}^{1}\!Y^e_I   \frac{\partial}{\partial l^e_1} \, {}^{1}\!Y^{e'}_{I'}  \right)
            \frac{\partial^2 (\tilde S_0 + \tilde S_1)}{ \partial l^{e'}_1\partial l^{e''}_{n'} }  \, .  \q\q
\ea
The first term on the right hand side vanishes for the same reason as before: the term inside the bracket is zero on flat configurations and the entire expression is a derivative in flat direction of this term. Concerning the second term, note that
\ba\label{dd6}
\sum_{n'=0,1,2}    \frac{\partial^2 (\tilde S_0 + \tilde S_1)}{ \partial l^{e'}_1\partial l^{e''}_{n'} }  {}^{(1)}\!y^{e''}_{n'}
&=&    \!\!\!\!   \!\!\! \sum_{n'=0,1,2}    \frac{\partial^2 (\tilde S_0 + \tilde S_1)}{ \partial l^{e'}_1\partial l^{e''}_{n'} } \, {}^{n'}\! T^{e''}_{I''}   {}^{(1)} \!y^{I''}_{n'}  +   \sum_{n'=0,1,2}    \frac{\partial^2 (\tilde S_0 + \tilde S_1)}{ \partial l^{e'}_1\partial l^{e''}_{n'} }  {}^{n'}\! T^{e''}_{\alpha''}    {}^{(1)} \! y^{\alpha''}_{n'}   \q\q\q
\ea
are the first order equations of motion associated to the edge $e'$ with the first term on the right hand side vanishing automatically.  Hence, $(I_1 I'_{1}  \Delta''_{n''} )$ vanishes if the first order equations of motion for the ${}^{(1)}\!y^\alpha_1$ are satisfied.
~\\

With {\bf (a)} and {\bf (b)} the remaining terms in $\mathfrak S_I$ are then given by
\ba\label{dd8}
\mathfrak{S}_I &\underset{\text{~}}{=} &(I_1\alpha'_0 I''_1)+ (I_1I'_1\alpha_0'')+( I_1I'_1\alpha''_1)+ (I_1\alpha'_1I''_1)+( I_1I'_1\alpha''_2) +(I_1\alpha'_2I''_1)
\nn\\
&&+ \sum_{n',n''=0,1,2}(I_1 \alpha'_{n'} \alpha''_{n''})  \nn\\
&\underset{\text{1. order  eom}}{=} &\q\sum_{n',n''=0,1,2}(I_1 \alpha'_{n'} \alpha''_{n''})  \q .
\ea
The first line of (\ref{dd8}) can be rearranged according to (\ref{dd5}) and (\ref{dd6}) to yield terms proportional to the first order equations of motion for the variables ${}^{(1)}y^\alpha_1$. If these first order equations are satisfied we therefore only remain with terms without any dependence on the first order gauge variables and without any second order (gauge and gauge invariant) variables. 
~\\
~\\
{\bf (c)} We will consider the second order of Hamilton's principal function -- that is the action evaluated on the solution -- and its derivative with respect to the background gauge parameter. \\
~\\
The action $\tilde S=\tilde S_0+ \tilde S_1$ expanded to second order reads
\ba\label{hope1}
\tilde S &=&  {}^{(0)}\tilde S + \varepsilon \sum_{n=0,2} \frac{\partial \tilde S}{\partial l^e_n} \,\, {}^{n}\!T^e_\Gamma \,
\left(  {}^{(1)}\!y^\Gamma_n + \varepsilon\,  {}^{(2)}\!y^\Gamma_n \right) +
\varepsilon \frac{\partial \tilde S}{\partial l^e_1} \,\, {}^{1}\!T^e_\Gamma \,
\left(  {}^{(1)}\!y^\Gamma_1+ \varepsilon\,  {}^{(2)}\!y^\Gamma_1 \right) \nn\\
&&+\q\q
\tfrac{1}{2} \varepsilon^2 \sum_{n',n''}  \frac{\partial^2 \tilde S }{\partial l^{e'}_{n'} \partial l^{e''}_{n''} }
\,\, {}^{n'}\!T^{e'}_{\Gamma'}  {}^{n''}\!T^{e''}_{\Gamma''} \,\,  {}^{(1)}\!y^{\Gamma'}_{n'} \, {}^{(1)}\!y^{\Gamma''}_{n''} \,\,\,\,+ \,\,\,\,\,\,O(\varepsilon^3) \q .
\ea
The zeroth order term does not depend on any (background) variables at time step $n=1$, since only extrinsic curvature angles appear in it (which can be expressed using length variables from only the boundary or time steps $n=0,2$). The same holds for the second term in the first line. The last term in the first line vanishes because of the zeroth order equations of motion and also its derivative  ${}^{1}\!Y^e_I \frac{\partial}{\partial l^e_1}$ vanishes even though we would like to solve for the first and second order variables. We remain with the second order terms. Using, as in (\ref{dd7}), the notation
\ba\label{hope2}
( \Delta'_{n'} \Delta''_{n''}):=\tfrac{1}{2}\sum_{\Delta',\Delta''}   \frac{\partial^2 \tilde S}{ \partial l^{e'}_{n'}  \partial l^{e''}_{n''}} 
    ,\, {}^{n'}\!T^{e'}_{\Delta'}     \,\,   {}^{n''}\!T^{e''}_{\Delta''}   \,\,   \varepsilon^2 
        \, {}^{(1)}\!y^{\Delta'}_{n'}\, \, {}^{(1)}\!y^{\Delta''}_{n''}  \q ,
\ea
we can analyze the terms according to their type. Firstly, notice that all terms with $n'=0$ and $n''=2$ or vice versa vanish. Secondly, all terms of the type $(I'_1 \Delta''_{n''})$ and $( \Delta'_{n'}I''_1)$ vanish as ${}^{1}T_I^e={}^{1}Y_I^e$ is a null vector of the Hessian of the action. These term still vanish if we apply another derivative ${}^{1}\!Y^e_I \frac{\partial}{\partial l^e_1}$ corresponding to infinitesimally changing the vertex at $n=1$ in the embedding flat space time, where the Hessian contracted with ${}^{1}\!Y^e_I$ identically vanishes. The same holds for terms of the type $(I'_{n'} \alpha''_{n''})$ which vanish either because of (\ref{ob2}) or (\ref{need1}).

We are left with the following second order terms
\ba\label{hope3}
{}^{(2)}\tilde S&=& (I'_0I''_0)+ (I'_2 I''_2) + \sum_{n',n''=0,1,2} (\alpha'_{n'}\alpha''_{n''}) \q .
\ea
The first two terms disappear under the action of a derivative ${}^{1}\!Y^e_I \frac{\partial}{\partial l^e_1}$ as is shown in {\bf (a)}.  For the other terms we obtain
\ba\label{hope4}
{}^{1}\!Y^e_I \frac{\partial}{\partial l^e_1}\sum_{n',n''=0,1,2} (\alpha'_{n'}\alpha''_{n''})    &=&
 \sum_{n',n''=0,1,2} (I_1 \alpha'_{n'}\alpha''_{n''}) + \mathfrak{E} \q .
\ea
The additional terms summarized as $\mathfrak{E}$ arise through the derivative ${}^{1}\!Y^e_I \frac{\partial}{\partial l^e_1}$ acting on the solutions for ${}^{(1)}y^{\alpha'}_1,{}^{(1)}y^{\alpha''}_1$ and on the components ${}^1T^{e'}_{\alpha'},{}^1T^{e''}_{\alpha''}$ . (We have to replace the variables ${}^{(1)}y^{\alpha'}_1,{}^{(1)}y^{\alpha''}_1$ by the solutions to obtain Hamilton's principal function. Also note that the derivatives with respect to the length $l^e_1$ are not acting on the components ${}^{n'}T^{e'}_{\alpha'}$ for $n'\neq 1$. The reason is that the expression ${}^{n'}Y^e_I  N^{n'}_{ee'}$ only involves background variables from time step $n'$. Similarly, the conditions (\ref{ob2}) on the matrix ${}^{n'}T^e_\Gamma$ only involve background variables from time step $n'$, hence one can also choose ${}^{n'}T^e_\Gamma$ to be of this type.) These terms $\mathfrak{E}$ are proportional to the first order equations of motion, however, and therefore vanish
\ba\label{hope5}
\mathfrak{E}&=& \tfrac{1}{2} \varepsilon^2 \sum_{n'=0,1,2} 
 \frac{\partial^2 \tilde S }{\partial l^{e'}_{n'} \partial l^{e''}_{1} }
\,\, {}^{n'}\!T^{e'}_{\alpha'}  \,  {}^{1}\!T^{e''}_{\alpha''} \,\,  {}^{(1)}\!y^{\alpha'}_{n'} \,\,\left( {}^{1}\!Y^e_I \frac{\partial}{\partial l^e_1} \,     {}^{(1)}\!y^{\alpha''}_{1} \right)
+
\nn\\
&& \tfrac{1}{2} \varepsilon^2 \sum_{n''=0,1,2} 
 \frac{\partial^2 \tilde S }{\partial l^{e'}_{1} \partial l^{e''}_{n''} }
\,\, {}^{1}\!T^{e'}_{\alpha'}   \, {}^{n''}\!T^{e''}_{\alpha''} \,\, \left( {}^{1}\!Y^e_I \frac{\partial}{\partial l^e_1}  {}^{(1)}\!y^{\alpha'}_{1} \right)\,   {}^{(1)}\!y^{\alpha''}_{n''}  
 + 
\nn\\
 && \tfrac{1}{2} \varepsilon^2 \sum_{n'=0,1,2} 
 \frac{\partial^2 \tilde S }{\partial l^{e'}_{n'} \partial l^{e''}_{1} }
\,\, {}^{n'}\!T^{e'}_{\alpha'} 
\left(  {}^{1}\!Y^e_I \frac{\partial}{\partial l^e_1} \,    {}^{1}\!T^{e''}_{\alpha''}    \right)  \,\,  {}^{(1)}\!y^{\alpha'}_{n'} \,\,   {}^{(1)}\!y^{\alpha''}_{1} 
+
\nn\\
&& \tfrac{1}{2} \varepsilon^2 \sum_{n''=0,1,2} 
 \frac{\partial^2 \tilde S }{\partial l^{e'}_{1} \partial l^{e''}_{n''} }
\,\,
\left(      {}^{1}\!Y^e_I \frac{\partial}{\partial l^e_1}         \,  {}^{1}\!T^{e'}_{\alpha'}            \right)  {}^{n''}\!T^{e''}_{\alpha''} \,\, {}^{(1)}\!y^{\alpha'}_{1} \,\,   {}^{(1)}\!y^{\alpha''}_{n''}    \nn\\
&\underset{\text{1. order  eom}}{=} & 0 \, .
\ea

As a consequence, we finally obtain
\ba\label{hope6}
\mathfrak{S}_I 
&\underset{\text{1. order  eom}}{=} & 
{}^{1}\!Y^e_I \frac{\partial}{\partial l^e_1} \,\, {}^{(2)}\tilde S \q .
\ea
This finishes the proof.\\~\\

%%%%%%%%%%%%%%%%%%%%%%%%%%%%%%%%%%%%%

To summarize, for the first non--linear order of the equations of motion (\ref{dd2}) the following situation arises:  the equations for $\Gamma=\alpha$ have to be used to determine the second order gauge invariant observables ${}^{(2)}\!y^\alpha$, as these only appear there. For the remaining equations of motions $\Gamma=I$, which contain only first order gauge invariant variables (if the first order equations of motion are satisfied), we do not have any variables left to solve for and we seem to have an inconsistent theory.

However, the remaining terms in the equations of motion for $\Gamma=I$ will generically depend on the background gauge parameters, which in a sense are zeroth order variables. Indeed, these equations of motion now have a precise interpretation, namely as equations which fix the background gauge such that the second order of Hamilton's principal function (which can also be called the effective action) depends minimally on this gauge.

This also entails that one can obtain a consistent expansion to higher order only for certain choices of the gauge parameter in the background solution.  For other choices one cannot expand the fluctuation variables $y^e$ in a power series in $\varepsilon$: the solutions corresponding to the gauge degrees of freedom have a lowest order term proportional to $\varepsilon^{-1}$, which can be interpreted as a change of the background gauge.  On the other hand, at the lowest non--linear order we find that the first and second order gauge variables $y^I$ remain undetermined. For the next order, i.e.\ an expansion of the action to fourth order, we expect that the equation of motion determine the first order gauge variables.\\

These considerations can be tested with the parametrized harmonic oscillator (and unharmonic generalizations). The action for one time step is given by
\ba\label{dd9}
S_n=\frac{1}{2}\frac{(q_{n+1}-q_{n})^2}{(t_{n+1}-t_n)} - \frac{1}{8} \omega ( q_{n}+ q_{n+1})^2 (t_{n+1}-t_n)\q .
\ea
We consider the variation of the variables at time step $n=1$ with fixed data at time steps $n=1,2$ and expand the action using $q_n=\varepsilon\,\, {}^{(1)}q_k+\varepsilon^2 \,\,{}^{(2)}q_n$ and  $t_n= {}^{(0)}\!t_n+\varepsilon {}^{(1)}t_n+\varepsilon^2 {}^{(2)}t_n$ to third order around the configurations $q_0,q_1,q_2=0$ and arbitrary $t_k$.  These configurations are solutions to the equations of motion with $t_k$ being the background gauge parameters. One finds that the second order equation of motion corresponding to $t_1$ is satisfied automatically only for $t_1=\tfrac{1}{2}(t_0+t_2)$. As a result, the higher order terms determine the time discretization. 

Moreover, if one just defines $q_k=0+\varepsilon y_k$ and $t_k={}^{(0)}\!t_k+\varepsilon z_k$ and takes the expansion of the action to third order as a definition of the dynamics one finds that the solution of $z$ is not analytic in $\varepsilon$. The lowest order rather scales with $\varepsilon^{-1}$, that is effectively changes ${}^{(0)}\!t_1$.
\\

In the canonical framework we can define the momenta at time step $n=1$ via the action $\tilde S_0$ and the action $\tilde S_1$. The contraction of these momenta with the null vectors ${}^{1}\!Y^e_I$ resulted in constraints. From the previous discussion we can conclude that the second order momenta (defined via $\tilde S_1$)  contracted with ${}^{(1)}\!Y^e_I$ are of the form
\ba\label{dd10}
{}^{1}\!Y^e_I\,\,\,{}^{(2)}\pi_e^1 &=&
- \frac{\partial^2 \tilde S_1}{\partial l^e_1 \partial l^{e'}_1}  \,\,{}^1\!T^e_I\,\, {}^1\!T^{e'}_{I'} \,\,   \,{}^{(2)}\!y^{I'}_1 \nn\\
  &&- \sum_{n''=1,2}             
      \frac{\partial^3  \tilde S_1}{\partial l^e_1 \partial l^{e'}_1   \partial l^{e''}_{n''}} 
       \,\,{}^1\!T^e_I\,\, {}^1\!T^{e'}_{I'}     \,\,   {}^{n''}\!T^{e''}_{\alpha''}   \,\,  
        \, {}^{(1)}\!y^{I'}_1\, \, {}^{(1)}\!y^{\alpha''}_{n''}   \nn\\
        &&- \q\q\q \tfrac{1}{2}       
      \frac{\partial^3  \tilde S_1}{\partial l^e_1 \partial l^{e'}_1   \partial l^{e''}_{1}} 
       \,\,{}^1\!T^e_I\,\, {}^1\!T^{e'}_{I'}     \,\,   {}^{1}\!T^{e''}_{I''}   \,\,   
        \, {}^{(1)}\!y^{I'}_1\, \, {}^{(1)}\!y^{I''}_{1}  \nn\\
     &&   - \tfrac{1}{2}\!\!\!\!\sum_{n',n''=1,2}             
      \frac{\partial^3  \tilde S_1}{\partial l^e_1 \partial l^{e'}_n   \partial l^{e''}_{n''}} 
       \,\,{}^1\!T^e_I\,\, {}^{n'}\!T^{e'}_{\alpha'}     \,\,   {}^{n''}\!T^{e''}_{\alpha''}   \,\, 
        \, {}^{(1)}\!y^{\alpha'}_{n'}\, \, {}^{(1)}\!y^{\alpha''}_{n''}   \q .
\ea
Note that only gauge variables ${}^{(1)}\!y^I,\,\, {}^{(2)}\!y^I$ from time step $n=1$ appear. The only variables from time step $n=2$ are the first order gauge invariant ${}^{(1)}\!y^\alpha_2$. Using the first order equations of motion (\ref{dyn2}), however, these can be expressed as linear combinations of variables ${}^{(1)}\!y^\alpha_1,\,\,{}^{(1)}\!\pi_\alpha^1$ at time step $n=1$. 

Hence, if we consider only the fluctuation variables ${}^{(k)}\!y,\,\,{}^{(k)}\!\pi$ with $k\geq 1$ as true variables we can also obtain at second order relations which only involve the variables at one time step. From this point of view one can still speak of constraints. However, these constraints are not automatically preserved by time evolution anymore. The reason is that the corresponding covariant equations (\ref{dd8}), which are exactly the condition for the preservation of the constraints, are not automatically satisfied. 

If we also consider the (gauge) parameters of the background solution as zeroth order variables, however, the second order terms of the constraints (\ref{dd10}) will in generic cases depend on these variables from time step $n=2$. In this sense the second order constraints are pseudo constraints. Not all the terms in (\ref{dd10}) will depend on the background gauge parameters at time step $n=2$ -- one can show that all terms with first or higher order gauge variables only depend on the background variables at time $n=1$. These are exactly the terms which cancel automatically in the covariant equations of motion (\ref{dd8}).

Using a similar rewriting as in equation (\ref{dd5}) and the first order equations of motion, the second order part of the constraints can be written as
\ba\label{dd11}
{}^{(2)}C_I&=&{}^{1}\!Y^e_I\,\,\,{}^{(2)}\pi_e^1
+ \frac{\partial^2 \tilde S_1}{\partial l^e_1 \partial l^{e'}_1}  \,\,{}^1\!T^e_I\,\, {}^1\!T^{e'}_{I'} \,\,   \,{}^{(2)}\!y^{I'}_1
\nn\\&&
 + \left( \,{}^{1}\!Y^e_I \frac{\partial }{\partial l^e_1} \,{}^{1}\!Y^{e'}_{I'}   \right) ({}^{1}\!T^{-1})^{\alpha''}_{e'} \, \, {}^{(1)}y^{I'}_1
\,\, {}^{(1)}\pi_{\alpha''}^1  %\nn\\&& 
-  \left( \,{}^{1}\!Y^e_I \frac{\partial }{\partial l^e_1} \,{}^{1}\!T^{e''}_{\alpha''}   \right)  \frac{\partial^2 \tilde S_1}{\partial l^{e'}_1 \partial l^{e''}_1}  \,\,{}^{1}Y^{e'}_{I'} \, \, {}^{(1)}y^{I'}_1 \,\, {}^{(1)}y^{\alpha''}_1 
 \nn\\
&& + \q\q\q 
\tfrac{1}{2}       
      \frac{\partial^3  \tilde S_1}{\partial l^e_1 \partial l^{e'}_1   \partial l^{e''}_{1}} 
       \,\,{}^1\!T^e_I\,\, {}^1\!T^{e'}_{I'}     \,\,   {}^{1}\!T^{e'}_{I''}   \,\,   
        \, {}^{(1)}\!y^{I'}_1\, \, {}^{(1)}\!y^{I''}_{1}  \nn\\
&& + \tfrac{1}{2}\!\!\!\!\sum_{n',n''=1,2}             
      \frac{\partial^3  \tilde S_1}{\partial l^e_1 \partial l^{e'}_n   \partial l^{e''}_{n''}} 
       \,\,{}^1\!T^e_I\,\, {}^{n'}\!T^{e'}_{\alpha'}     \,\,   {}^{n''}\!T^{e'}_{\alpha''}   \,\, 
        \, {}^{(1)}\!y^{\alpha'}_{n'}\, \, {}^{(1)}\!y^{\alpha''}_{n''}   \q ,
\ea
  where ${}^{(1)}\!y^\alpha_2$ appearing in the last line can be substituted by an expression involving only variables at $n=1$
  \ba
  {}^{1}\!T^e_\alpha \frac{\partial^2 \tilde S_1}{\partial l^e_1 \partial l^{e'}_2} \, {}^{2}\!T^{e'}_{\alpha'}\,\,\, {}^{(1)}\!y^{\alpha'}_2 
  &=& -\pi^1_\alpha - {}^{1}\!T^e_\alpha  \frac{\partial^2 \tilde S_1}{\partial l^e_1 \partial l^{e'}_1} \, {}^{1}\!T^{e'}_{\alpha'}  \,\,{}^{(1)}\!y^{\alpha'}_1 \q .
  \ea
    
Note that  the constraints might remain exact constraints, i.e.\ relations between variables (including zero order variables) from only one time step, to higher or even all orders. The latter is the case for tent moves at  four--valent vertices, which lead to flat dynamics. The full non-linear constraints for this situation are given by (\ref{flata}). \\

The phenomenon that not all solutions of the linearized theory can be completed to solutions of  the full theory is similar to  the occurrence of linearization instabilities in continuum general relativity for space times with compact spatial slices \cite{fischer}. In our case this phenomenon occurs because the solutions of the full theory are unique (if there are no flat vertices), whereas the linearized solutions admit firstly freedom for the choice of the background gauge and secondly freedom in the choice of the first--order gauge parameters. The consistency conditions eliminate this gauge freedom order by order. There is an important difference to the linearization instabilities appearing in \cite{fischer}: whereas there the additional conditions are on the first order (physical) modes, the consistency conditions here fix the zeroth--order background gauge variables (and presumably the higher order equations fix the higher order gauge variables).

\subsection{Example: Symmetry reduced five valent vertex}

%Maybe calculate numerically the pseudo constraints, similar to the plots in section 6 of paper 1 (Figure 8)? Can we do that with the data we already have (we just need to solve the t-equation for this)\\
%Apply to symmetry reduced 5-valent vertex?

Here we will consider the consistency equation arising from the second order equations of motion for the symmetry reduced five valent vertex, described in section (\ref{beispiel1}). The variables we have to deal with are the lengths $a_n,b_n$ for $n=0,1,2$. We fix data at $n=0,2$ and consider the equations of motion with respect to $a_1,b_1$. 

To begin with, we have to (numerically) find solutions for the lengths of the tent pole $t_0,t_1$. This can be done to second order in an expansion around the flat configuration determined by the initial values ${}^{(0)}\!a_0=1,{}^{(0)}\!b_0=1$ and
\ba\label{hope01}
{}^{(0)}\!t_0=\tfrac{1}{10} +\tau \, ,\q\q {}^{(0)}\!t_1=\tfrac{1}{10}-\tau
\ea
where $\tau$ is the background gauge parameter, determining the position of the vertex at $n=1$ in the background flat space time. Note that  ${}^{(0)}\!a_2, {}^{(0)}\!b_2$, the background data at $n=2$ are independent of $\tau$. 

This way, we can obtain the effective action $\tilde S$ expanded to third order.  From this action we can obtain the first and second order equations of motion for ${}^{(1)}y^a_1,{}^{(1)}y^b_1,{}^{(2)}y^a_1,{}^{(2)}y^b_1$. Solving the first order equation corresponding to the derivative with respect to $a_1$ for  ${}^{(1)}y^a_1$ and using the solution in the other first order equation of motion, one will find that it is automatically satisfied. This is the signature of the exact gauge freedom for the linearized theory. 

Using the first order solution we can solve the second order equation corresponding to the derivative with respect to $a_1$ for  ${}^{(2)}y^a_1$. Again, we use this solution in the other second order equation corresponding to the derivative with respect to $b_1$. For instance, for $\tau=0$ we find that we have to solve the equation (ignoring terms of order $10^{-8}$ arising due to numerical errors)
\ba\label{hope02}
0 &=& -0.5464921\, ({}^{(1)}\!y_0^a)^2 
- 0.9715415 \, ({}^{(1)}\!y_0^b)^2           + 1.4573123  \, ({}^{(1)}\!y_0^a)({}^{(1)}\!y_0^b)  \nn\\&&  
+ 1.0936310\,   ({}^{(1)}\!y_0^a) ({}^{(1)}\!y_2^a)
 - 
    1.3269178  \,({}^{(1)}\!y_0^a) ({}^{(1)}\!y_2^b) 
  \nn\\&&- 1.4581747 \,({}^{(1)}\!y_0^b)  ({}^{(1)}\!y_2^a)  + 1.7692237\, ({}^{(1)}\!y_0^b)({}^{(1)}\!y_2^b)
    \nn\\&&
 - 0.5471391\, ({}^{(1)}\!y_2^a)^2- 0.8054603\, ({}^{(1)}\!y_2^b)^2  +1.3277030 \,({}^{(1)}\!y_2^a) ({}^{(1)}\!y_2^b)  \q .
\ea
However, as expected from the previous discussion, all variables at time step $n=1$ have dropped out. We thus have to find a value for the background gauge parameter $\tau$ such that the remaining second order equation is satisfied. A priori one would expect that the value of $\tau$ has to depend on the boundary data ${}^{(1)}\!y^a_n,{}^{(1)}\!y^b_n$ with $n=0,2$. But (as for the parametrized harmonic oscillator) it turns out, that this equation can be solved independently from these first order boundary data.  All coefficients in (\ref{hope02}) which are non--vanishing for $\tau=0$ vanish simultaneously  (within numerical accuracy) for $\tau=-0.008303982$. We conjecture that there is a general mechanism which ensures that the second order terms $\mathfrak{S}_I$ can be made to vanish independently of the value of the first order boundary data.

\section{Conclusion and discussion}

In this work we introduced a canonical formalism for discretized gravity which exactly reproduces the dynamics as defined by the discrete action. In this way also the exact and approximate symmetries of the action are reproduced as first class constraints and pseudo constraints, respectively. For linearized Regge calculus, which exhibits exact gauge symmetries, we obtained Abelian constraints valid for arbitrary triangulated Cauchy surfaces. The momenta and constraints are local  functions (as opposed to the suggestion in \cite{friedmann}) and have an immediate geometric interpretation as generators of vertex displacements. The constraints can be shown to commute and to be preserved by the tent move dynamics as defined by the quadratic order of the action. 
Although this can be proved by staying entirely in the canonical framework, it is much easier to remember that the constraints follow from the symmetries of the action. For instance, the constraints are Abelian because second derivatives (of the action) commute.

%One might argue that flat triangulations are just a special class of solutions and the existence of symmetries for these solutions are rather an exception. We want to emphasize, however, that the continuum limit is controlled by the ratio of curvature scale to the discretization scale, that is the deficit angles. In the continuum limit these deficit angles should become very small. The near flat space behaviour of Regge calculus is therefore important for the continuum limit and the existence of exact gauge symmetries for the linearized theory is important for obtaining the correct number of propagating physical degrees of freedom. 

Discretized constraints are often derived by performing first a canonical analysis of the continuum action and then discretizing the resulting constraints. This often leads to a change of the constraint algebra from first to second class \cite{loll, friedmann}, i.e.\ the constraints close only modulo terms proportional to some power in the lattice spacing. This work suggests to not only qualify discretized constraints under this criterium but to also consider the properties of the constraints near solutions relevant for the continuum limit -- here the flat solutions.

We argued that the constraints to the first non--linear order will aquire some dependence on the background gauge parameters and turn therefore into pseudo constraints \cite{gambini}. These pseudo constraints are not automatically preserved under tent move evolution anymore. Rather some of the equations of motion turn into consistency conditions, selecting a specific background gauge and ensuring that the constraints are now preserved under time evolution.

These consistency conditions can be rewritten as the derivatives with respect to the background gauge parameters of the quadratic part of Hamilton's principal function. Note that this quadratic part can be computed in the linearized theory.  Despite the fact that the linearized theory displays exact constraints and gauge symmetries (in the fluctuation variables), the associated Hamilton's principal function is not independent of the background gauge, or in other words the discretization. These findings entail that a consistent perturbative expansion is only supported for specific background choices. For other choices solutions are not analytic in the expansion parameter.

Although the background gauge is fixed by the lowest non--linear order dynamics the first and second order gauge variables remain undetermined. We expect that these are fixed one after the other in the higher order dynamics, so that in the $n$-th order dynamics the $n$th and $(n-1)$th order of the gauge variables remain undetermined.

%%%%%%%%%%%%%%%%%%%%%%%%%%%%%%%%%%%%%%%%%%%%%%%

We want to underline that although we found pseudo constraints starting with second order for Regge calculus, there might  be alternative discretization schemes which lead to first class discrete constraints for the full theory. In fact, as was shown in \cite{bahrdittrich09b} for Regge calculus with a cosmological constant, where the symmetries are broken, it is possible to find an alternative discretization of the action with exact symmetries also for curved solutions. The methods used in \cite{bahrdittrich09b} are the so--called perfect actions, which exactly reproduce the dynamics (and therefore symmetries) of the continuum. These perfect actions can be obtained by a  coarse graining procedure (from the continuum). Obtaining the perfect action for 4d gravity will be very complicated, but might be achievable in a perturbative approach \cite{ta2}. The considerations in this paper give necessary prerequisites towards this end.

If one is only interested in a particular class of solutions, for instance almost homogeneous solutions for cosmological applications, one might try to derive improved actions by adapting the discretization to the chosen background. In the case of cosmology, it would be interesting to have an action which displays gauge symmetries for homogeneous solutions and in this way ensure the correct number of propagating degrees of freedom on cosmological backgrounds.  See also  \cite{bojocosmo} for related discussions in loop quantum cosmology and \cite{carlocosmo} for a proposal for using tent moves in a (quantum) cosmological setting.

We hope that these results can be useful for connecting covariant and canonical approaches to quantum gravity, in particular spin foam models and canonical loop quantum gravity. This has been achieved in 3d \cite{perez} (where the symmetries are exact for the full non--linear theory) but is an outstanding problem in 4d \cite{noui}. Spin foam models can be understood as partition functions for a discretized theory \cite{spinfoam}. In canonical loop quantum gravity the central ingredient defining the dynamics is given by the quantum constraints \cite{qsd}. 

This work suggests a derivation of a canonical quantum theory from the covariant one, that is a canonical theory directly derived from the amplitudes associated to spin foam models. As spin foam models are based on a discretization of the Plebanski action one expects the symmetries to be broken. Hence, one can doubt whether on the discrete level, i.e.\ before taking any continuum limit or sum over triangulations, (exact) quantum constraints exist for the full theory that would reproduce the dynamics defined by the spin foam model. Nevertheless, it should be possible to define a canonical dynamics with discrete time that reproduces the spin foam amplitudes. To this end it would be useful to perform a canonical analysis of the discretized Plebanski action, a task that was only recently achieved for the continuum action in \cite{sergej}. In particular, in \cite{dittrichryan} one can find the relation between the canonical phase spaces of discretized Plebanski theory and of Regge calculus. Also in \cite{dittrichryan} it was pointed out that there exists a class of triangulations which admits only flat solutions and on which the symmetries are therefore exact. For this class of triangulations it might therefore be possible to derive quantum constraints from the quantum amplitudes. To accomplish this, the symmetries of the amplitudes have to be better understood. Interesting results in this direction and a connection to the quantum constraints can be found in \cite{barrett,valentin}.

\vspace{2cm}

\begin{center}{\bf  ACKNOWLEDGMENTS}\end{center}
~\\
BD would like to thank Benjamin Bahr and Valentin Bonzom for interesting discussions and express her gratitude for an invitation to the CPT Marseille and fruitful discussions with the quantum gravity group there, especially Carlo Rovelli and Simone Speziale. PAH acknowledges the support of the German Academic Exchange Service (DAAD) through a doctoral research grant and is grateful for a travel grant of Universiteit Utrecht.

%%%%%%%%%%%%%%%%%%%%%%%%%%%%%%%%%%%%%%%%%%%%%%%%%%%%%%%%%%%%%%%%%%%%%%%%%%%%%%%%%%%%%%%%%%%%%%%%%%%%%%%%%%%%%%%%%%%%%%%%%%%%%%%%%%%%%%%%%%%%%%%%%%%%%%%%%%%%%%%%%%%%%%%%%%%%%%%%%%%%%%%%%%%%%%%%%%%%%%%%%%%%%%%%%%%%%%%%%%%%%%%%%%%%%%%%%%%%%%%%

\end{document}